\documentclass[%
 reprint,
 aps,
 prl,
]{revtex4-1}

\usepackage{amsmath}
\usepackage{amssymb}
\usepackage{amsthm}
\usepackage{array}
\usepackage{graphicx}
\usepackage{float}
\usepackage{booktabs}
\usepackage[caption=false]{subfig}

\usepackage{natbib,mathtools}
\usepackage{psfrag}
\usepackage{mathrsfs}

\newcommand{\rmb}{\mathrm{b}} 
\newcommand{\rmc}{\mathrm{c}} 
 
\newcommand{\rme}{\mathrm{e}}

\newcommand{\rmi}{\mathrm{i}}

\newcommand{\rmr}{\mathrm{r}}

\newcommand{\rmx}{\mathrm{x}}  
\newcommand{\rmy}{\mathrm{y}}  
\newcommand{\rmz}{\mathrm{z}}  
\newcommand{\rmB}{\mathrm{B}}

\newcommand{\rmM}{\mathrm{M}}

\newcommand{\rmX}{\mathrm{X}}  
\newcommand{\rmY}{\mathrm{Y}}

\newcommand{\bfe}{\mathbf{e}}

\newcommand{\bfk}{\mathbf{k}}

\newcommand{\bfx}{\mathbf{x}}

\newcommand{\bfB}{\mathbf{B}} 
 
\newcommand{\bfD}{\mathbf{D}} 
\newcommand{\bfE}{\mathbf{E}}

\newcommand{\bfH}{\mathbf{H}}

\newcommand{\bfK}{\mathbf{K}} 
 
\newcommand{\bfM}{\mathbf{M}}

\newcommand{\bfQ}{\mathbf{Q}} 
\newcommand{\bfR}{\mathbf{R}}

\newcommand{\bfX}{\mathbf{X}}

\newcommand{\overbar}[1]{\mkern 1.5mu\overline{\mkern-1.5mu#1\mkern-1.5mu}\mkern 1.5mu}

\newcommand{\eff}{\mathrm{eff}}

\newcommand{\pp}{{\prime\prime}}  

\newcommand{\scatcoeff}{N}  
\newcommand{\omegand}{\omega_\rmB}  
\newcommand{\omegad}{\omega}

\begin{document}

\preprint{APS/123-QED}

\title{Violating the Energy-Momentum Proportionality of Photonic Crystals in the Low-Frequency Limit} 

\author{Michael J. A. Smith}
 \affiliation{School of Mathematics, The University of Manchester,  Manchester M13 9PL, United Kingdom} 
 \email{michael.j.smith@manchester.ac.uk}
 
\author{Parry Y. Chen}
\affiliation{Unit of Electro-optic Engineering, Faculty of Engineering Sciences, Ben-Gurion University, Beer Sheva, Israel}
\affiliation{
 School of Physics and Astronomy, Raymond and Beverly Sackler Faculty of Exact Sciences, Tel Aviv University, Tel Aviv, Israel
}

\date{\today} 

\begin{abstract} \noindent

We  theoretically show that  the  frequency and momentum of a photon are not necessarily proportional to one another at low frequencies in     photonic crystals  comprising materials with     positive- and negative-valued   material properties.    We rigorously determine    closed-form  conditions   for  the light cone  to emanate from points other than the origin of $k$ space, ultimately decoupling the first band from the origin and demonstrating light propagation at zero energy with nonzero crystal momentum.   We also numerically show that  first bands   can   originate from an arbitrary Bloch coordinate as well as from multiple   coordinates simultaneously.  
\end{abstract}

\maketitle

When a photon propagates through a dielectric medium at low frequencies, it satisfies the energy-momentum ($E$-$k$) relation $E = c \hbar |\bfk|$, where $c$ is the phase velocity in the medium \cite{joannopoulos2011photonic,born1964principles}. This relation ensures that at zero energy, the photon possesses zero momentum. Fundamental relations of this type are prevalent throughout nature and are not isolated to photons, for example, electrons propagate through a crystal lattice as $E = \hbar^2 /(2 m_\mathrm{eff}) |\bfk|^2$ at low energies, where $m_\mathrm{eff}$ is the effective mass \cite{kittel2005introduction}.   This proportionality is fundamental for the study of particles and fields in relativistic mechanics, particle physics, and quantum mechanics.

\begin{figure*}[t]

\makebox[ \textwidth]{
     \subfloat[\label{fig:conventionalGamma1}]{%
 
        \includegraphics[width=0.3225\textwidth]{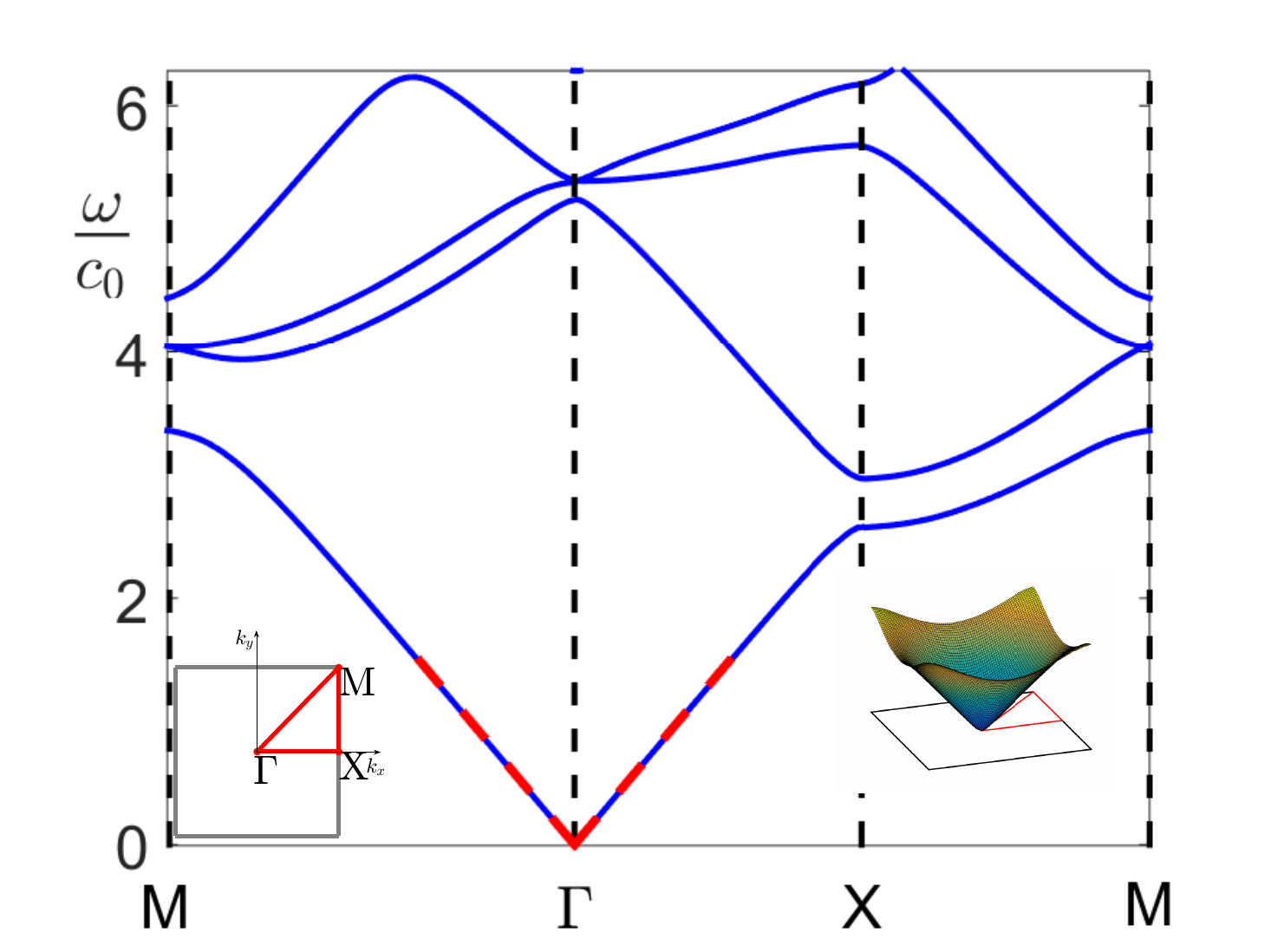}
     }
      
     \subfloat[\label{fig:Mptex1}]{%
       \includegraphics[width= 0.3225 \textwidth]{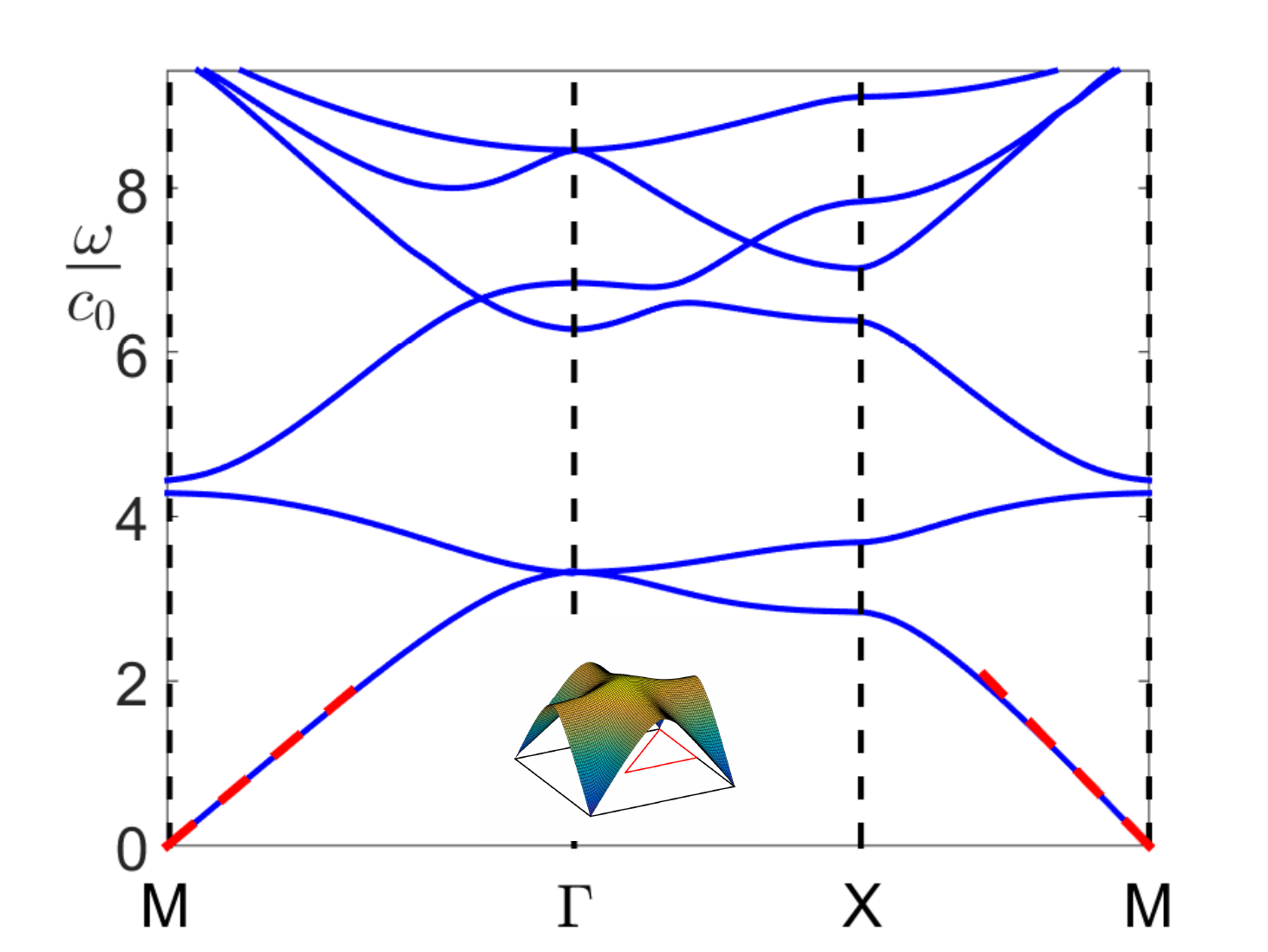}
     }
        
       \subfloat[\label{fig:Xpics1}]{%
       \includegraphics[width=0.3225 \textwidth]{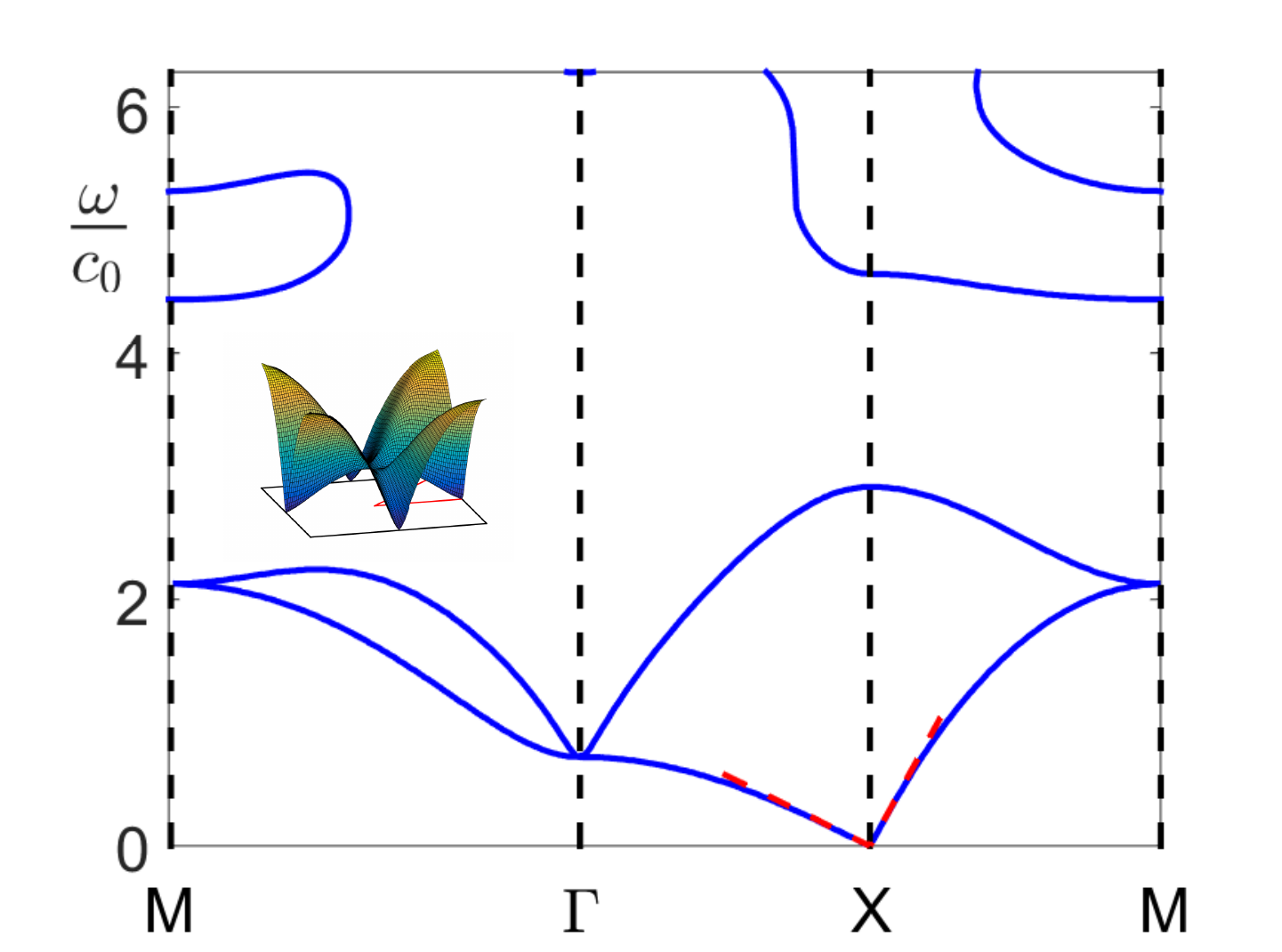}
     }
      }

      \caption{Band diagrams   for   square array of cylinders embedded in air     with    \protect\subref{fig:conventionalGamma1}      $\varepsilon_\rmc = 1$ and $\mu_\rmc = 2$, \protect\subref{fig:Mptex1}  $\varepsilon_\rmc = -1$ and $\mu_\rmc = 2$, and  \protect\subref{fig:Xpics1}  $\varepsilon_\rmc \approx -0.53$ and $\mu_\rmc \approx -10.22$.  Dashed red lines denote   low-frequency descriptions       \eqref{eq:disprelGamma},  \eqref{eq:kvk0}, and  \eqref{eq:disprelXellipse}, respectively.  All figures use  lattice period    $a=1$,    radius   $a^\prime = 0.3a$, and dipolar approximation.   Inset: first Brillouin zone with path parametrization $\Gamma$$X$$M$; first band surfaces over first Brillouin zone.\label{fig:conventionalGamma}    }
  
   \end{figure*}

In this Letter, we break the conventional  low-frequency $E$-$k$ proportionality for photons,   obtaining     relations of the form $E = C\, |\bfk - \Xi|$, where $\Xi$ denotes a high-symmetry point of the reciprocal lattice and $\bfk$ is   the crystal momentum (see Figs.~\ref{fig:conventionalGamma1}-c).
This is achieved in  two-dimensional  photonic crystals comprising materials with {\it positive-definite} and {\it negative-definite} \cite{smith2004metamaterials,veselago1968electrodynamics,pendry2000negative} optical properties.  We present explicit    conditions on the constituent properties and  explicit forms for the proportionality constants $C$,   for all high-symmetry coordinates of a   square lattice. 
Furthermore, we numerically demonstrate the existence of other novel low-frequency behaviors, including   photonic crystals with $E = C_1 \, |\bfk| + C_2 \,|\bfk - \bfX|$, where $C_j$ are proportionality constants and  $X$ is a high-symmetry point (see Fig.~\ref{fig:nonmagfigs5}). Such  unconventional behavior contrasts the   standard outcomes for light in photonic crystals, where  either $E$-$k$ proportionality is supported,  or there exists a complete   band gap \cite{fan1996large},  at low frequencies.

In the  nonstandard  photonic settings we describe,   massless photons are predicted to  propagate as massive polaritons which  travel superfluidically through the medium   \cite{lerario2017room}. Consequently, our findings have the potential to  motivate the  development of new photonic devices, and to deepen our understanding of light in structured  media. The behaviors   we describe     complement  existing observations in optical systems   incorporating negative-definite materials,  such as  folded band surfaces with infinite group velocities \cite{chen2011folded}, cloaking and superresolution  \cite{guenneau2007cloaking,helsing2011spectral}, and   new  types of band gaps \cite{li2003photonic}.   Analogies to our  low-frequency $E$-$k$ relations  may be found  in the electronic properties of transition-metal perovskites,  where the first   band is centered about   high symmetry points other than the origin \cite{harrison1989electronic,cora1997transition}.

We begin by considering the modes of  the time-harmonic form  of the source-free Maxwell equations in a   nondispersive and lossless system with  Bloch vectors   $\bfk = (k_{x}, k_{y},0)$.  This    wave vector restriction reduces Maxwell's equations to    the   Helmholtz equation
 \begin{equation}
 \label{eq:helmholtz}
\nabla_\perp \cdot \left( \varepsilon_\rmr^{-1} \nabla_\perp H_\rmz \right) +  \omegad^2 c_0^{-2}  \mu_\rmr H_\rmz = 0,
 \end{equation}
 for   fields polarized as   $\bfH = (0,0,H_\rmz)$. Here, $ \nabla_\perp \equiv(\partial_\rmx,\partial_\rmy)$, $\varepsilon_\rmr$   is the relative permittivity, $\mu_\rmr$ the relative permeability, $\omegad$   the angular   frequency, and $c_0$ is the speed of light in vacuum.   We consider an array of infinitely extending isotropic cylinders, periodically positioned in the $(x,y)$ plane at the coordinates of a square lattice, and which are embedded in an infinitely extending isotropic background material. In the background and cylinder domains,   material constants   are allowed to be negative-valued.  At the  cylinder edges  we impose      continuity conditions, and     between unit cells we impose   Bloch--Floquet conditions.  This admits the system \cite{movchan2002asymptotic}
 \begin{equation}
 \label{eq:dispeq}
 \scatcoeff_l  \mathcal{B}_l + \sum_{m=-\infty}^{\infty} (-1)^{l+m} S_{m-l}^\rmY  \mathcal{B}_m = \boldsymbol{0},
 \end{equation}
  where $S_m^\rmY = S_m^\rmY(\omegand,\bfk_\rmB;n_\rmb)$ denotes lattice sums (see   Supplemental Material \cite{suppmatref}\nocite{mcphedran1996low,poulton2000eigenvalue,mcphedran2000lattice,linton1998greens,twersky1961elementary,chin1994greens,bensoussan1978asymptotic,jikov2012homogenization,bergman1979dielectric,mcphedran1980electrostatic,busch1998photonic,linton2010lattice,abramowitz1964handbook,nicorovici1996analytical}),  $\omegand = \omegad/c_0$,     $\bfk_\rmB = (k_{x}, k_{y})$, $\mathcal{B}_m$ are     amplitudes of the cylindrical-harmonic basis functions, and $N_m$   are   inverse cylindrical-Mie coefficients. Where applicable,   subscripts $\rmb$ and $\rmc$ denote  the  background and cylinder properties, respectively.  The dispersion equation  for the crystal  is given by the vanishing determinant   of \eqref{eq:dispeq}, which we     truncate to dipolar order.

  We now outline the procedure for determining when a  low-frequency band surface     emerges from the $\Gamma$ point. First, we    evaluate   expansions  for  $N_m$ in $\omega_\rmB$ (these     are lengthy, see  Supplemental Material \cite{suppmatref}).  Next, we determine   closed-form expressions    for   the $S_l^\rmY$   in  \eqref{eq:dispeq} at low frequencies and about the $\Gamma$ point; these are obtained following \citet{chen2016evaluation} (also extensive, see Supplemental Material \cite{suppmatref}). Assuming that   $\omegand = \alpha k_\rmB$, where $\alpha$ is real and positive-valued, we subsequently obtain series coefficients for $S_l^\rmY$ in $\omegand$ alone.  Substituting  the  expansions for $N_m$ and $S_l^\rmY$ into \eqref{eq:dispeq}, the zero determinant condition is satisfied to the lowest order  
 %at  $O(\omegand^{-6})$ 
 for  $\alpha$ such that 
 \begin{equation}
\label{eq:disprelGamma}
\omegand = \left\{ 
\frac{1}{\varepsilon_\rmb}
\left( \frac{1+f  \tau }{ 1-f \tau}  \right)
\frac{   1 }{ \mu_\rmb+f    (\mu_\rmc-\mu_\rmb) } 
\right\}^{1/2} k_\rmB,
\end{equation}
 where $\tau = ( \varepsilon_\rmb -   \varepsilon_\rmc )/( \varepsilon_\rmb +   \varepsilon_\rmc )$, $f = \pi a^{\prime 2}/a^2$ is the filling fraction, $a^\prime$ denotes the radius of the cylinders, and $a$ is the lattice period. Thus,  the constituent permittivity and permeability values can be negative, but provided $\alpha>0$ then a band surface will emerge from $\Gamma$.   Here $\alpha = 1/n_\mathrm{eff}$ where $n_\mathrm{eff}$   is the effective refractive index \cite{movchan2002asymptotic}.

    \begin{figure*}[t]
  
\makebox[ \textwidth]{

     \subfloat[\label{fig:conventionalGamma1c}]{%
       \includegraphics[width=0.3225\textwidth]{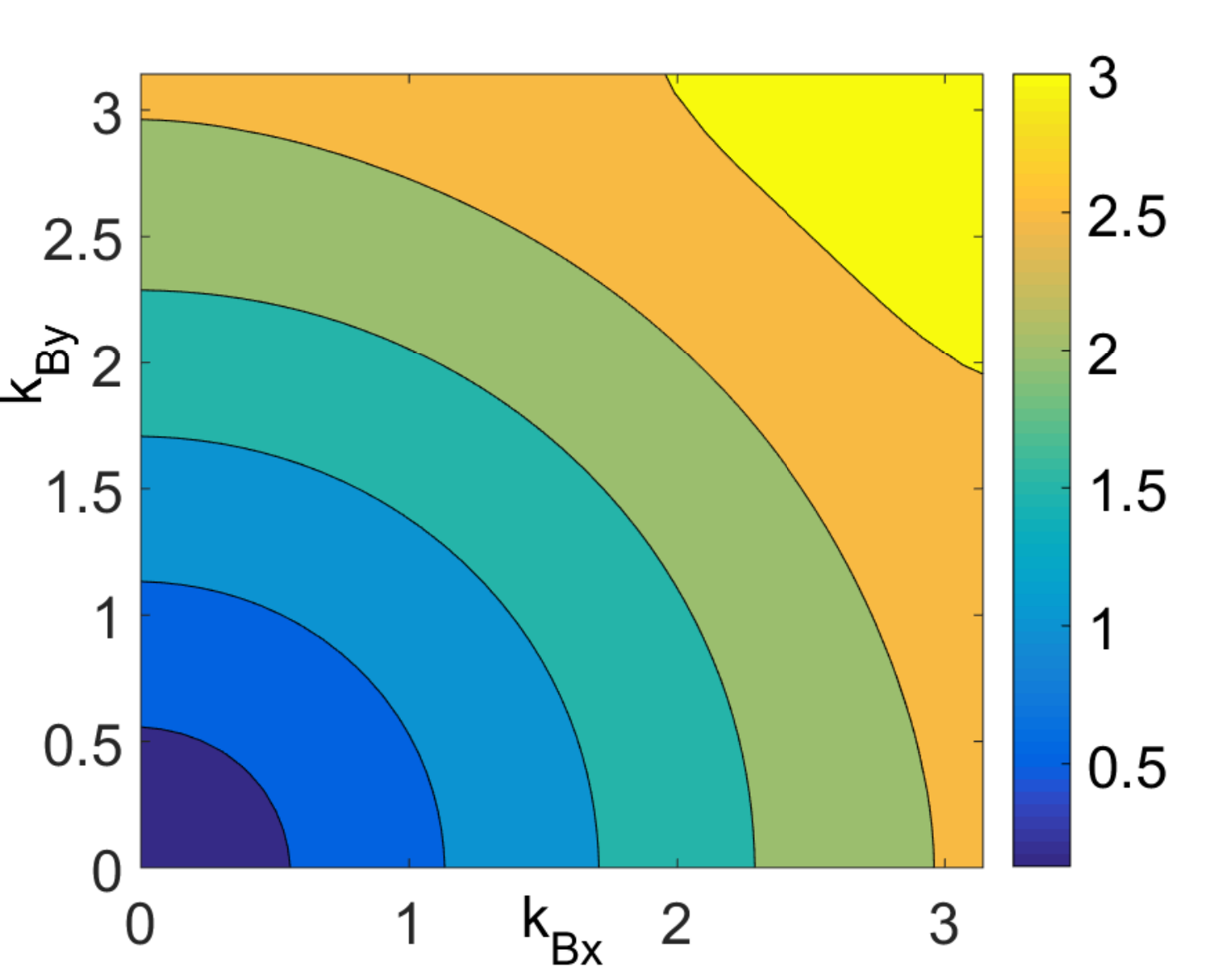}
     }
     
       \subfloat[\label{fig:Mptex1c}]{%
       \includegraphics[width= 0.3225 \textwidth]{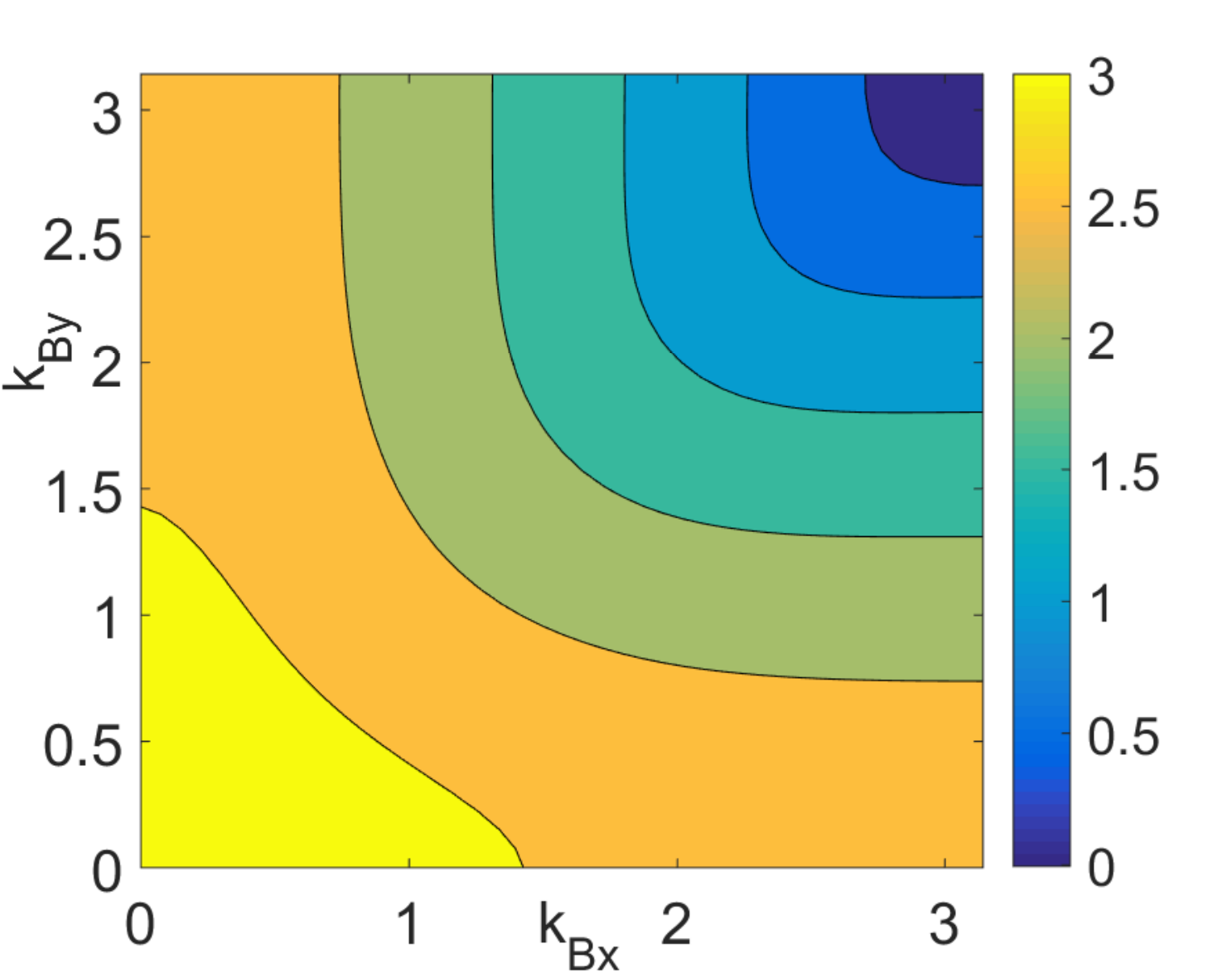}
     }
     
    \subfloat[\label{fig:Xpics1c}]{%
       \includegraphics[width=0.3225 \textwidth]{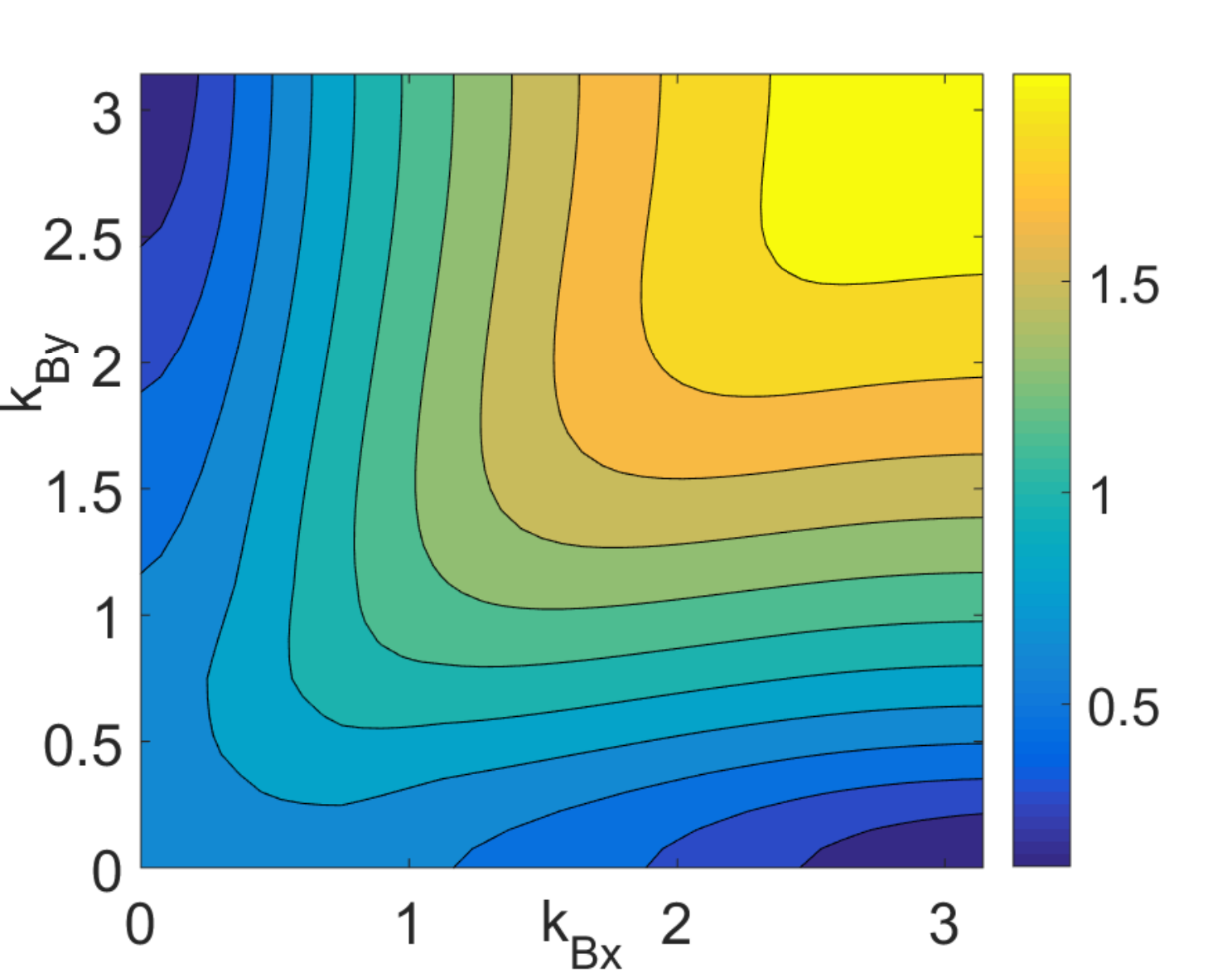}
     }
     
      }

     \caption{Isofrequency   contours      of    first band surface $\omega/c_0$    for configurations   in   Fig.~\ref{fig:conventionalGamma}.  \label{fig:allcontours}}
  
   \end{figure*}

   Next, we determine the conditions   and asymptotic behavior   of a   band that emerges from the $M$ point at low frequencies. As before, we derive  asymptotic forms for    the  $S_l^\rmY$  sums near $M$ (see   Supplemental Material \cite{suppmatref}) and assume  $\omegand = \alpha^\prime k_\rmB^\prime$, where $\alpha^\prime$ is real and positive-valued,  $\bfk_\rmB^\prime = \bfk_\rmB - \bfM$ and $\bfM = (\pi/a,\pi/a)$. This assumption yields series coefficients for  $S_l^\rmY$  in $\omegand$   (given in Supplemental Material \cite{suppmatref}).   Substituting  these expansions  for   $S_l^\rmY$ and $N_m$  into     \eqref{eq:dispeq},   the  zero determinant condition is satisfied to the lowest orders   when  
\begin{equation}  
\label{eq:epsM}
\varepsilon_\rmc = -\varepsilon_\rmb,
\end{equation}
 and for $\alpha^\prime$   such that
  \begin{multline}
 \label{eq:kvk0}
   \omegand  = \left\{ \frac{1}{  8 \pi^{ 2}  } \left| 64 \pi^4 \rme^{4 \rmi \theta_\rmB^\prime} + \Gamma(\tfrac{1}{4} )^8 \right|^{1/2} 
 \phantom{\left[ \varepsilon_\rmb \mu_\rmc + 2 \varepsilon_\rmb \mu_\rmb
        \log\left( \frac{16   \pi^2}{f   \Gamma(\tfrac{1}{4})^4}  \right) \right]^{-1/2}}
 \right. \\ 
 \left.  \times \left[ \varepsilon_\rmb \mu_\rmc + 2 \varepsilon_\rmb \mu_\rmb
        \log\left( \frac{16   \pi^2}{f   \Gamma(\tfrac{1}{4})^4}  \right) \right]^{-1/2}  \right\} k_\rmB^{\prime},
\end{multline}
where $\Gamma(z)$ is the Gamma function.    That is, a band surface is supported from   $M$   at   low frequencies   provided    \eqref{eq:epsM}    and  $\alpha^\prime>0$ 
  are satisfied.  The condition $\varepsilon_\rmc = -\varepsilon_\rmb$    corresponds to an {\it anomalous resonance}   in      quasistatic problems \cite{nicorovici1994optical,nicorovici2007quasistatic}   (discussed below).

Likewise,  for the $X$ point at low frequencies, having derived       asymptotic  forms for the   $S_l^\rmY$ sums near $X$ (see Supplemental Material \cite{suppmatref})  we assume that $\omegand = \alpha^{\prime \prime} k_\rmB^\pp$, where $\alpha^{\prime \prime}$ is real and positive-valued,  $\bfk_\rmB^\pp = \bfk_\rmB - \bfX$, and $\bfX = (\pi/a,0)$.     Substituting    the resulting   expansions   for   $S_l^\rmY$, and $N_m$,    into \eqref{eq:dispeq},    the   zero determinant condition is satisfied to leading order      when  
\begin{equation}
\label{eq:condCm2Em2}
\varepsilon_\rmc = \left( \frac{\zeta - 16   \pi^2 }{\zeta+ 16  \pi^2} \right) \varepsilon_\rmb,
\end{equation}
where $\zeta =   \Gamma(\tfrac{1}{4})^4  f$. At   the next order, provided
\begin{equation}
\label{eq:condepscX}
\mu_\rmc = \left( \frac{   \zeta ( \zeta - 64     \pi^2 )  
 + 512   \pi^4 \log\left( {\zeta}/{(32   \pi^2) }\right)   }{(\zeta - 16 \pi^2)^2}  \right)\mu_\rmb,
\end{equation}
  then we obtain  the low-frequency dispersion relation
\begin{subequations}
\begin{equation}
\label{eq:kk0X}
\omegand = \left\{  \frac{ \left( 16 \Gamma(\tfrac{1}{4})^4 \pi^2+ 64 \pi^4 \rme^{2\rmi \theta_\rmB^{\pp}} -\Gamma(\tfrac{1}{4})^8 \rme^{-2\rmi \theta_\rmB^{\pp}}   \right) }{16 \Gamma(\tfrac{1}{4})^4 \pi^2  \varepsilon_\rmb \mu_\rmb} \right\}^{1/2} k_\rmB^\pp.  
\end{equation}
However,   the slope  in \eqref{eq:kk0X} is only real-valued  along   $\Gamma X$ and $XM$. Numerical investigations confirm     elliptical contours at low frequencies; interpolating between these   paths with the ansatz     $\omegand^2 = \alpha_\rmx^\pp k_{\rmB \rmx}^{\pp 2} + \alpha_\rmy^\pp k_{\rmB \rmy}^{\pp 2}$   we   obtain  
 \begin{multline}
\label{eq:disprelXellipse}	 
\omegand^2 = \left( \frac{ 16 \Gamma(\tfrac{1}{4})^4 \pi^2 + 64 \pi^4 - \Gamma(\tfrac{1}{4})^8 }{16 \Gamma(\tfrac{1}{4})^4  \pi^2 \varepsilon_\rmb \mu_\rmb} \right) k_{\rmB \rmx}^{\pp 2}  \\
+ \left( \frac{ 16 \Gamma(\tfrac{1}{4})^4 \pi^2 - 64 \pi^4 + \Gamma(\tfrac{1}{4})^8 }{16 \Gamma(\tfrac{1}{4})^4  \pi^2 \varepsilon_\rmb \mu_\rmb} \right)  k_{\rmB \rmy}^{\pp 2},
\end{multline}
  for the first band surface as $\omegand \rightarrow 0$ and  as $\bfk_\rmB \rightarrow X$. 
\end{subequations}
Hence, a band surface is supported from $X$ at low frequencies provided  \eqref{eq:condCm2Em2},   \eqref{eq:condepscX},  and $\varepsilon_\rmb \mu_\rmb >0$   are satisfied.  In  \eqref{eq:condCm2Em2}, the proportionality factor  is      negative-valued for $ f\lesssim 0.914$, and the proportionality factor in  \eqref{eq:condepscX} is negative-valued for all $f$, demonstrating that highly restrictive sign-changing conditions must be satisfied in both $\varepsilon_\rmr$ and $\mu_\rmr$ so that  the first band   emerges from $X$.  The number of conditions for each high-symmetry Bloch coordinate is entirely due to the different asymptotic behaviors of   $S_l^\rmY$.   For arbitrary Bloch coordinate origin, we anticipate that the number of conditions  will change significantly.
  
     \begin{figure*}[t]
 
\makebox[ \textwidth]{
     \subfloat[\label{fig:nonmagfigs1}]{%
       \includegraphics[width=0.3225\textwidth]{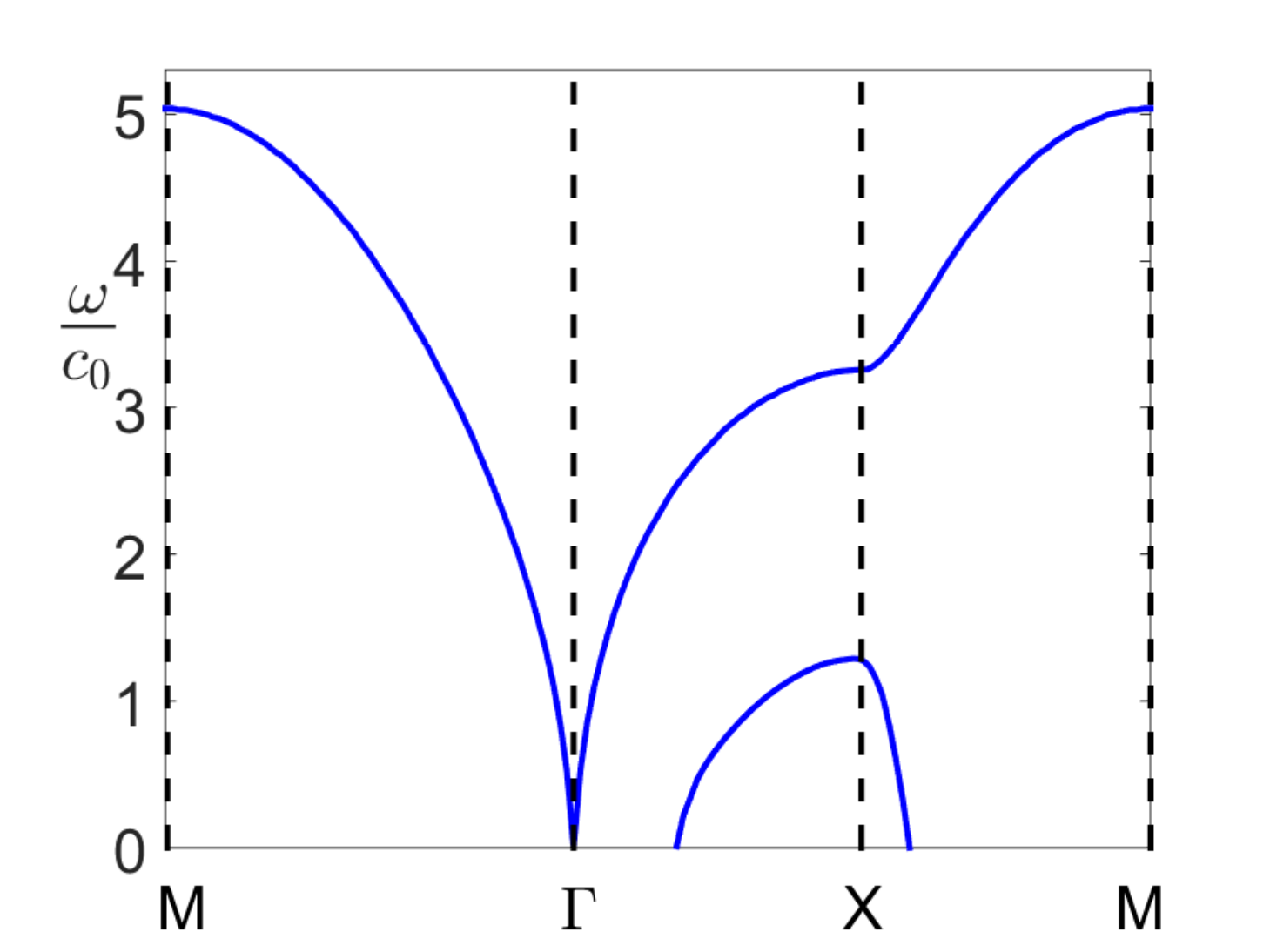}
     }
      
     \subfloat[\label{fig:nonmagfigs2}]{%
       \includegraphics[width=0.3225\textwidth]{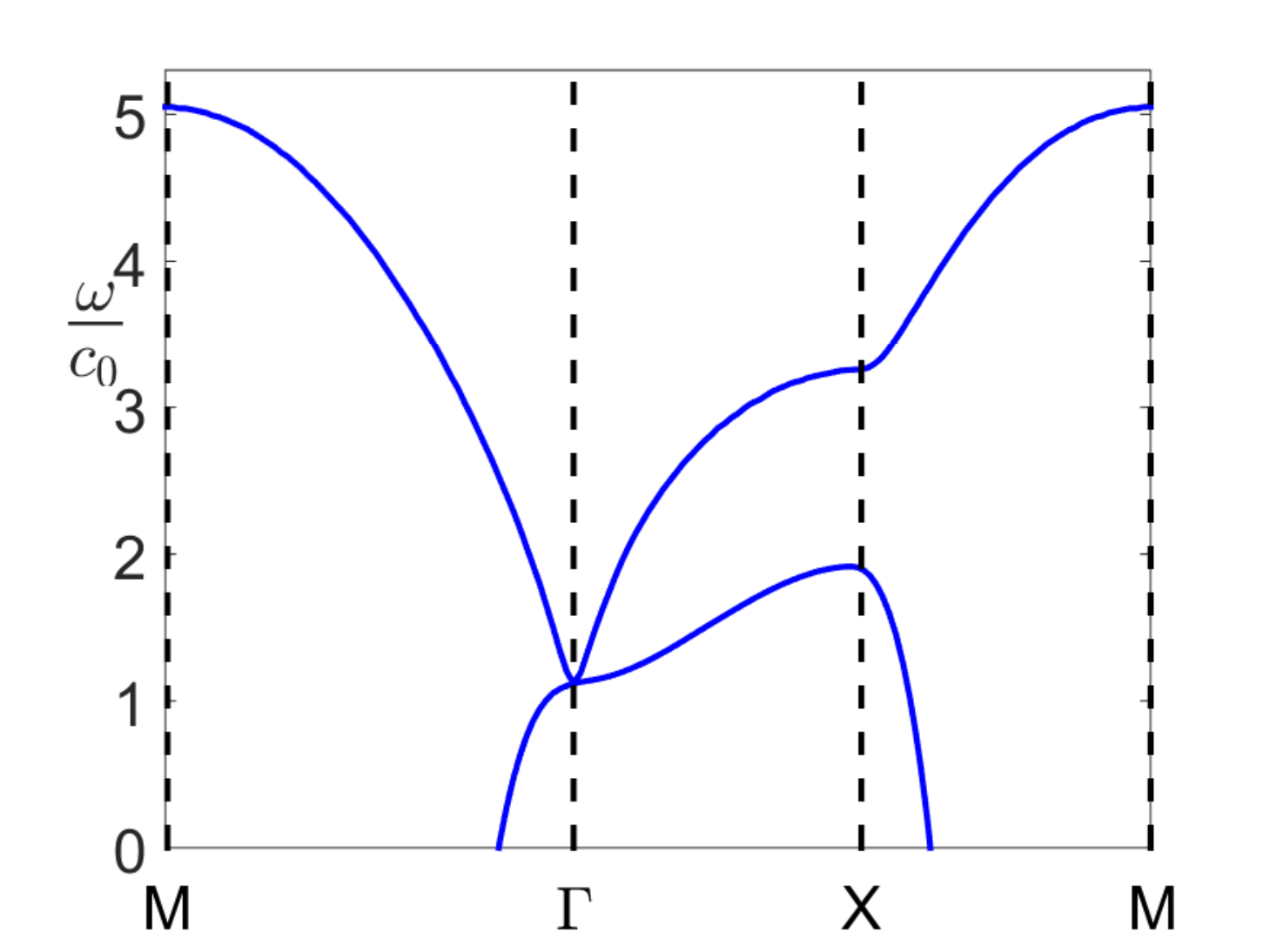}
     }
        
     \subfloat[\label{fig:nonmagfigs3}]{%
       \includegraphics[width=0.3225\textwidth]{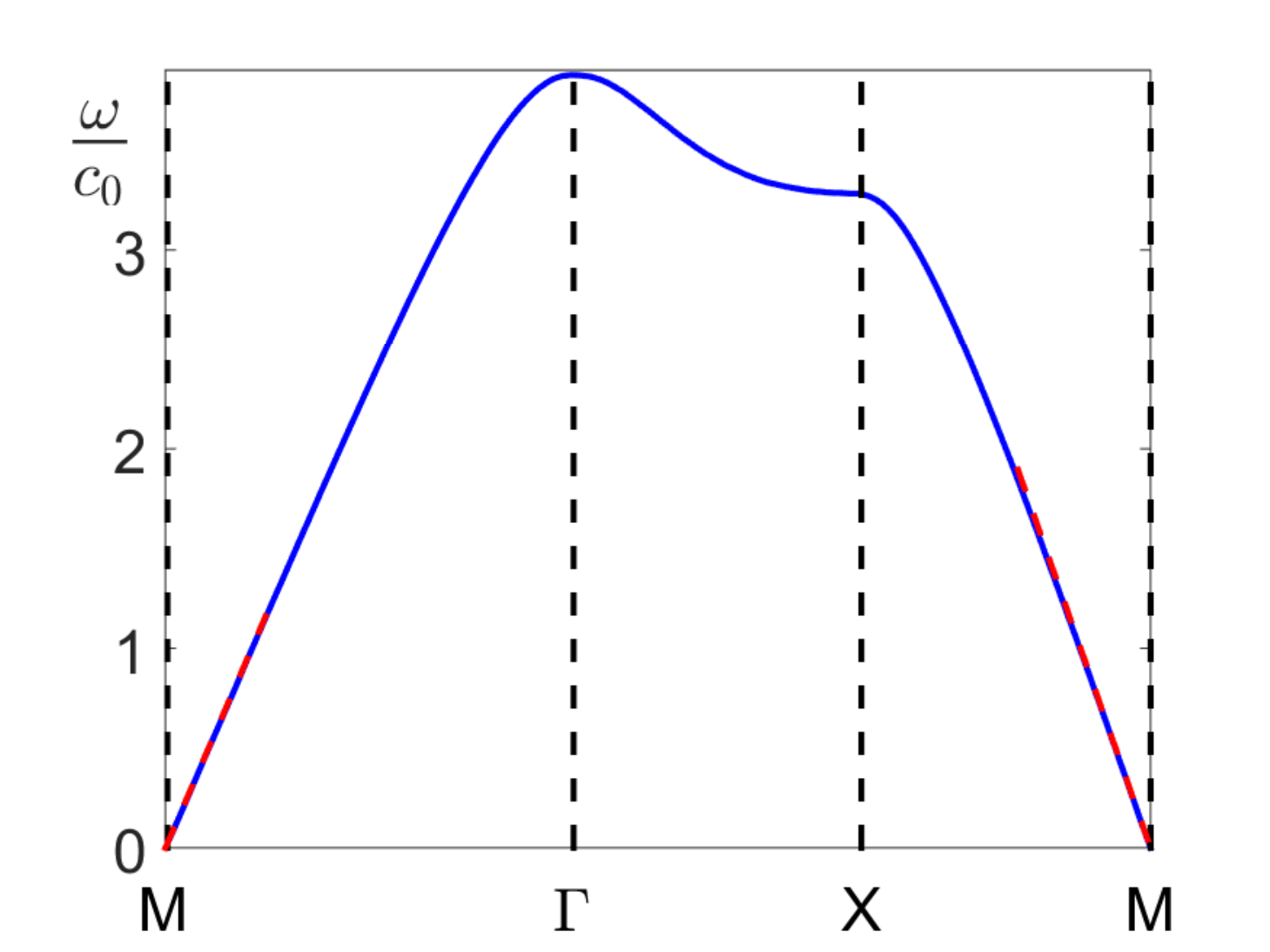}
     }
      }

\makebox[ \textwidth]{
  \subfloat[\label{fig:nonmagfigs4}]{%
       \includegraphics[width=0.3225\textwidth]{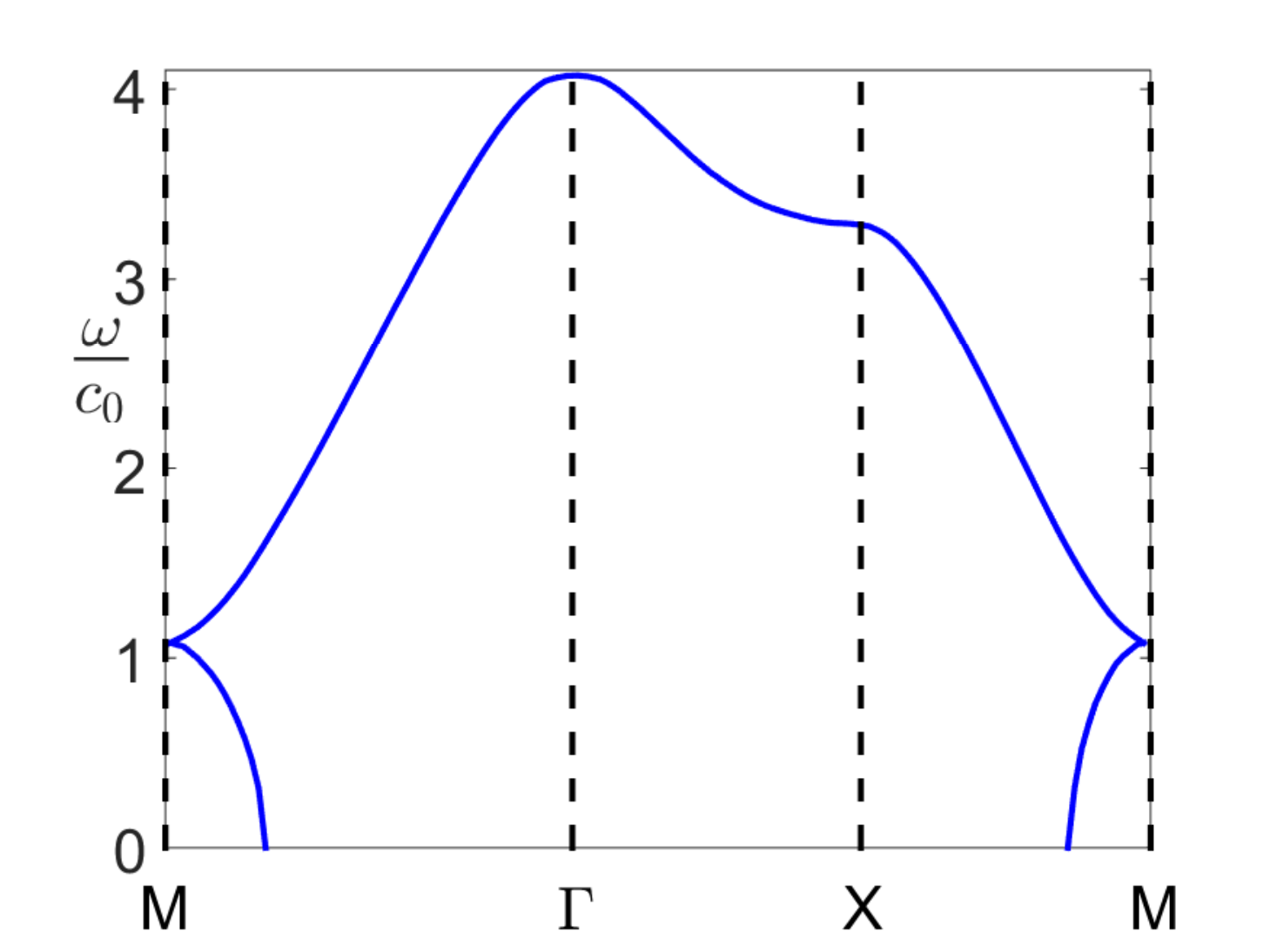}
     }
      
     \subfloat[\label{fig:nonmagfigs5}]{%
       \includegraphics[width=0.3225\textwidth]{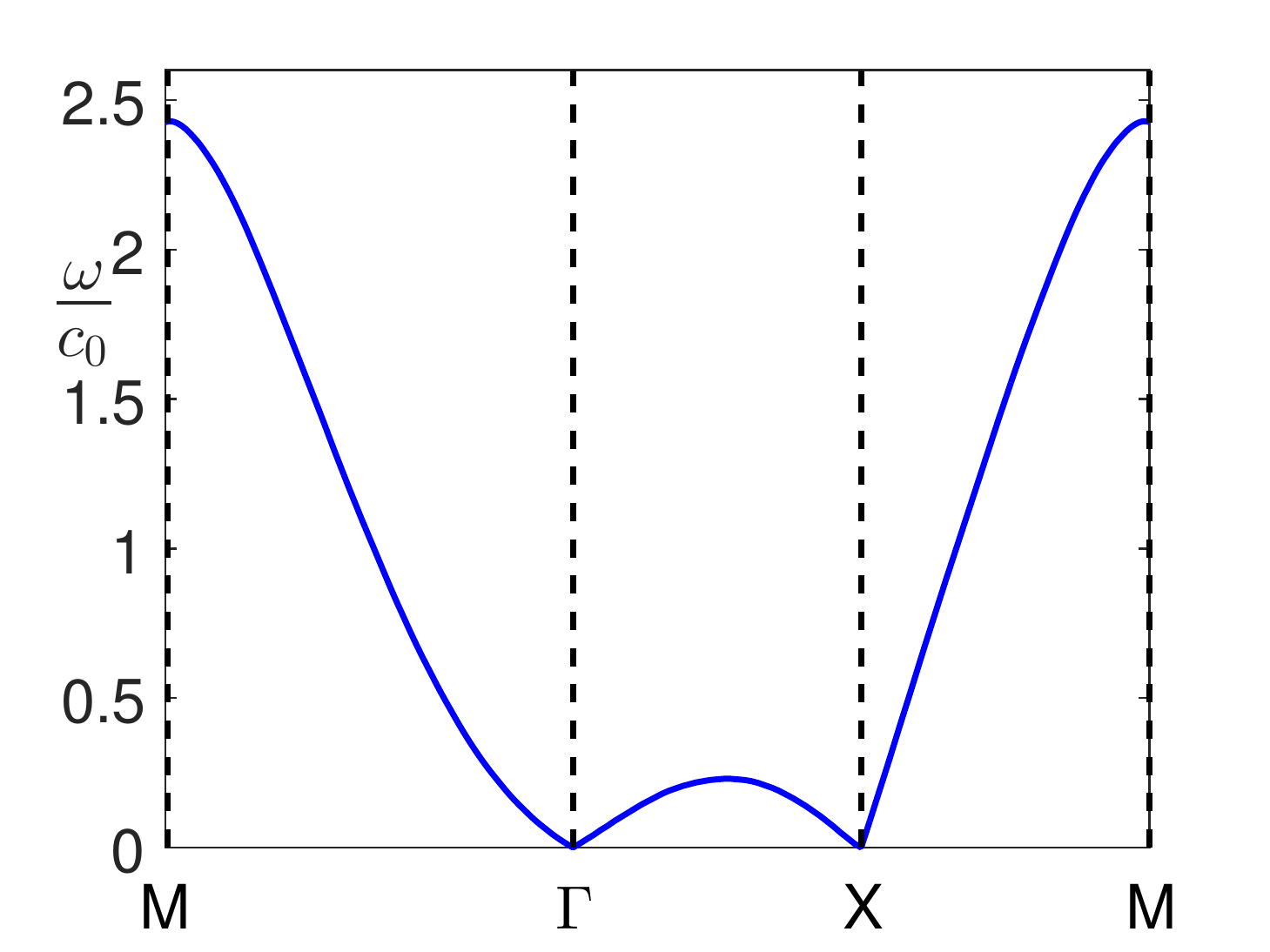}
     }
        
     \subfloat[\label{fig:nonmagfigs6}]{%
       \includegraphics[width=0.3225\textwidth]{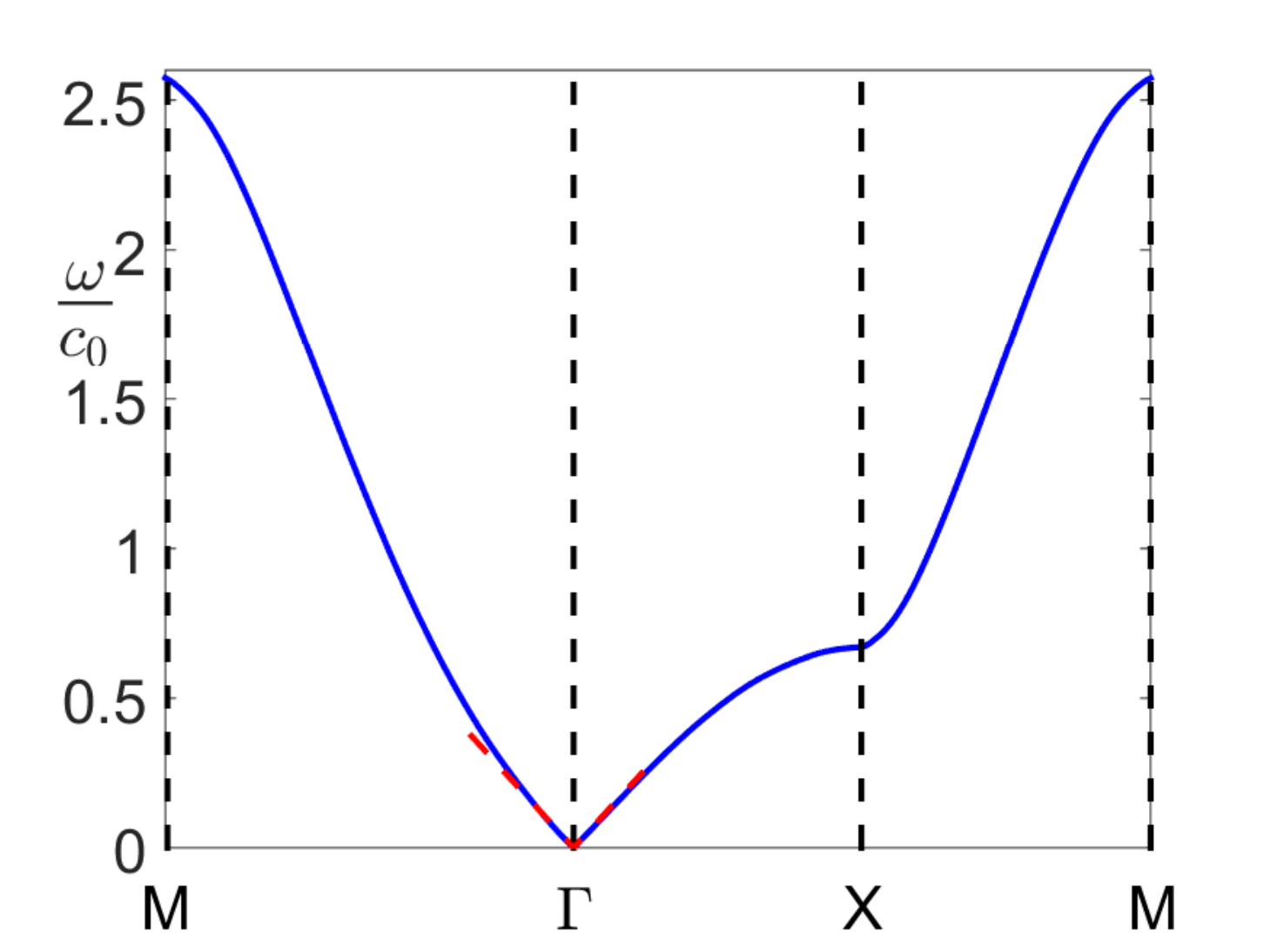}
     }
}

         \caption{Band diagrams for  square array of nonmagnetic cylinders ($\mu_\rmc = 1$)  embedded in   air   ($\varepsilon_\rmb = \mu_\rmb = 1$) with    \protect\subref{fig:nonmagfigs1}  $\varepsilon_\rmc = -0.55$, \protect\subref{fig:nonmagfigs2}  $\varepsilon_\rmc = -0.58$,  \protect\subref{fig:nonmagfigs3}  $\varepsilon_\rmc = -1$,    \protect\subref{fig:nonmagfigs4} $\varepsilon_\rmc = -1.1$, \protect\subref{fig:nonmagfigs5}  $\varepsilon_\rmc \approx - 1.896$, and     \protect\subref{fig:nonmagfigs6} $\varepsilon_\rmc = -2.1$. Dashed red lines in \protect\subref{fig:nonmagfigs3} and \protect\subref{fig:nonmagfigs6} are     approximations \eqref{eq:kvk0}   and \eqref{eq:disprelGamma}, respectively.   All figures use a dipolar approximation, lattice period    $a=1$, and radius      $a^\prime = 0.3a$.   \label{fig:nonmagfigs}}
   \end{figure*}

We now compare the  asymptotic forms  above  against results from a fully numerical treatment of \eqref{eq:dispeq} within the dipole truncation.    We begin by validating   \eqref{eq:disprelGamma} for a regular photonic crystal; in Fig.~\ref{fig:conventionalGamma1} we present the band diagram of a  representative   crystal  with    $\varepsilon_\rmc = 1$ and $\mu_\rmc = 2$ embedded in   air ($\varepsilon_\rmb = \mu_\rmc = 1$). As expected, the first band   emanates from  the $\Gamma$ point, and  \eqref{eq:disprelGamma} shows excellent agreement.  In Fig.~\ref{fig:Mptex1} we consider  $\varepsilon_\rmc = -1$ and $\mu_\rmc = 2$, where  the first band emanates from the $M$ point as described by \eqref{eq:kvk0} at low frequencies, also with excellent agreement.   In Fig.~\ref{fig:Xpics1} we consider  $\varepsilon_\rmc \approx -0.53$ and $\mu_\rmc \approx -10.22$  satisfying \eqref{eq:condCm2Em2} and  \eqref{eq:condepscX}. Here, the  slope differs along   $\Gamma X$ and $X M$, demonstrating   twofold symmetry  as $\omegand\rightarrow 0$. The   asymptotic estimate  \eqref{eq:kk0X}    shows  excellent agreement near $X$ at low frequencies. The   diagram also possesses folded bands \cite{chen2011folded} at high frequencies.

 In Fig.~\ref{fig:allcontours} we present     iso-frequency contours for the first band surfaces of    the   photonic crystals considered in Fig.~\ref{fig:conventionalGamma}, over a quarter   of the first Brillouin zone. In Fig.~\ref{fig:conventionalGamma1c} we see  a cone  ($\infty$-symmetric) as $\omegand\rightarrow 0$,   whereas in Fig.~\ref{fig:Mptex1c} we observe fourfold symmetric contours, as expected from the $\mathrm{exp}(4\rmi \theta_\rmB^\prime)$ dependence in \eqref{eq:kvk0}.   In Fig.~\ref{fig:Xpics1c} we see that the first    band  has twofold symmetric contours,   as expected from \eqref{eq:disprelXellipse}. These low-frequency symmetries contrast with the electronic band diagrams of graphene (and photonic analogues to graphene)  where the high-energy band emanates from    the $K$ point   and about the Fermi energy $E_F$ as an ideal cone    \cite{reich2002tight,zhang2005experimental,zandbergen2010experimental}.

Having   numerically validated the  $E$-$k$ relations \eqref{eq:disprelGamma}, \eqref{eq:kvk0}, and \eqref{eq:kk0X}, we now briefly demonstrate that magnetic constituents are not necessary to observe exotic low-frequency behaviors. Low-frequency descriptions of nonmagnetic photonic crystals   emanating from $\Gamma$  and $M$ are obtained by the replacements $\mu_\rmb, \mu_\rmc \mapsto 1$ in  \eqref{eq:disprelGamma} and \eqref{eq:kvk0}    above. This is despite the fact that the $N_m$ coefficients for  nonmagnetic crystals exhibit  different leading order behavior for small $\omegand$ (see Supplemental Material \cite{suppmatref}).  However,   the new leading-order behavior of $N_m$   yields a nonmagnetic analogue to {\eqref{eq:condepscX}  of the form
\begin{equation}
\label{eq:Xnonmagmu}
\zeta + 8\pi^2 - 16 \pi^2 \log \left( \zeta/(32 \pi^2) \right) = 0,
\end{equation}
 which is not satisfied for any $f$. As such, low-frequency emanation from   $X$   as   $\omegand = \alpha^{\prime \prime} k_\rmB^\pp$ or  $\omegand^2 = \alpha_\rmx^\pp k_{\rmB \rmx}^{\pp 2} + \alpha_\rmy^\pp k_{\rmB \rmy}^{\pp 2}$ is not supported for nonmagnetic crystals.

 In Fig.~\ref{fig:nonmagfigs} we present first band(s)    for   a selection of crystals  comprising  nonmagnetic cylinders ($\mu_\rmc =1$) in   air, and describe their evolution as $\varepsilon_\rmc$ is varied from $0 < \varepsilon_\rmc \leq -2.1$.  For values   $0 > \varepsilon_\rmc \gtrsim -0.5274$   \eqref{eq:condCm2Em2}, a  single first band   emanates from the $\Gamma$ point,   analogously to the $\Gamma$-emerging band in Fig.~\ref{fig:nonmagfigs1}. At  $\varepsilon_\rmc \approx -0.5274$,   a band emerges from the $X$ point, giving rise to two first band surfaces at  low frequencies, as in Fig.~\ref{fig:nonmagfigs1}. This $X$-emergent surface eventually  forms  a double degeneracy at the $\Gamma$ point at $\varepsilon_\rmc \approx -0.56$, and thereafter, as shown in Fig.~\ref{fig:nonmagfigs2},   becomes the first band surface,  pushing the existing $\Gamma$ centered band to higher frequencies. As we proceed in $\varepsilon_\rmc$, the    origin of the first band travels   along the $\Gamma M$ and $\Gamma X$ symmetry planes to the $M$ point at $\varepsilon_\rmc = -1$ (the anomalous resonance condition), as shown in Fig.~\ref{fig:nonmagfigs3}. Thereafter, a new first band emerges at the $M$ point whose origins move along   high symmetry planes towards the $\Gamma$ point. At the same time, this emergent band pushes the existing $M$ centered band to higher frequencies. Both of these behaviors are demonstrated in Fig.~\ref{fig:nonmagfigs4}.   The new first band eventually takes the form given in Fig.~\ref{fig:nonmagfigs5} where it emanates from both the $\Gamma$ and $X$ points simultaneously at $\varepsilon_\rmc = (\zeta + 16   \pi^2)/(\zeta - 16   \pi^2) \approx -1.896$.  As we proceed in $\varepsilon_\rmc$, the first eigenfrequency at $X$ becomes nonzero and the first band    emanates from the $\Gamma$ point alone, as demonstrated by Fig.~\ref{fig:nonmagfigs6}. Undoubtedly further behaviors are observed beyond $\varepsilon_\rmb<-2.1$,  however, this falls  outside the scope of the present work.  Note that in Figs.~\ref{fig:nonmagfigs3}  and \ref{fig:nonmagfigs6}   the dipolar approximations \eqref{eq:kvk0}   and \eqref{eq:disprelGamma} are superposed, respectively,  and  show    excellent  agreement at low frequencies.  The   examples above, in consideration with \eqref{eq:Xnonmagmu}, emphasize that  the absence of a band from a single Bloch coordinate does not preclude the   emergence of   bands from multiple Bloch coordinates simultaneously; this observation has important implications for determining the existence of   band gaps at low frequencies. We emphasize that the band structure smoothly transitions between the examples shown above.

     In summary, we have     determined new low-frequency $E$-$k$ relations for photons in   2D  photonic crystals. These   relations, and the conditions for their existence, are given explicitly for  first bands with  origins at the $\Gamma$, $X$, and $M$ points of a square lattice. In general, we have found that    sign changes in the properties of the constituents are required for the first band to originate from coordinates away from $\Gamma$ at low frequencies. We have also demonstrated that photonic crystals can possess low-frequency $E$-$k$ relations   with origins   at one or more arbitrary Bloch coordinates.     Given that all conventional  photonic crystals   possess either a low-frequency band gap or a band surface emanating from   $\Gamma$, this work has significant implications for the homogenization   of periodic media,  theoretical descriptions of light propagation in complex media, as well as  future  photonic  crystal designs.    In the latter case, the closed-form conditions we obtain represent a powerful design tool for determining filling fractions 
and background materials for a given   cylinder material, and vice versa.  However, an important consequence of the $M$-point condition coinciding with   the anomalous resonance condition is that   the slope of the     $M$-emerging band \eqref{eq:kvk0} is not necessarily accurate beyond a dipolar truncation;  further investigations are    required to accurately determine   behaviors      near resonance (see Supplemental Material \cite{suppmatref}).     Preliminary results for   hexagonal lattices at the anomalous resonance reveal  a band surface emerging from  the $K$ point, implying  that     the first band   originates from the furthest edge of the irreducible Brillouin zone when on resonance.  Away from the anomalous resonance condition, we believe that experimental validation is feasible for the crystals we describe, as all   emergence conditions   (i.e., \eqref{eq:condCm2Em2} and \eqref{eq:condepscX})      are valid for complex-valued $\varepsilon_\rmb$ and $\mu_\rmb$. When $\varepsilon_\rmb$ and $\mu_\rmb$ possess     moderate loss,  and $\varepsilon_\rmc$ and $\mu_\rmc$ satisfy the necessary emergence conditions, we find    the   band diagrams to be unchanged  (see Supplemental Material \cite{suppmatref}).     A natural next step for this work is    1D and 3D photonic structures; 2D photonic crystals were only chosen  for analytical convenience. Finally, we emphasize that our approach  extends  readily to phononic and other   systems. 
 
M.J.A.S. acknowledges discussions with R.C. McPhedran, S. Guenneau, R.V. Craster, support from the ERC (279673), and support from the EPSRC (EP/L018039/1). 
 
    \clearpage

 \onecolumngrid
\section{Supplemental Material for ``Violating the energy-momentum proportionality of photonic crystals in the low-frequency limit''}

 \section{  Extended problem formulation} \label{sec:problemform}
 We begin by considering the time-harmonic form  of the source-free Maxwell equations, in a   non-dispersive and lossless system. The domain comprises a two-dimensional array of infinitely extending isotropic cylinders that are periodically positioned at the coordinates of a square lattice in the $(x,y)$ plane and embedded in an isotropic background material.    That is, we  examine the band structure of  the Maxwell  equations
 \begin{align}
 \label{seq:maxwelleqns} \nonumber
 \nabla \times \bfH = -\rmi \omegad  \bfD, &\quad  \nabla \times \bfE = \rmi \omegad   \bfB,\\ 
\nabla \cdot \bfD = 0,  &\quad \nabla \cdot \bfB = 0,
 \end{align}
 having assumed that fields possess the dependence  $\mathrm{exp}(-\rmi \omegad t)$. We then assume the linear constitutive relations
\begin{align}
\label{seq:constit}
\bfD = \varepsilon_0\varepsilon_\rmr \bfE, \quad \mbox{ and } \quad \bfB = \mu_0 \mu_\rmr \bfH,
\end{align}
where the relative permittivity and permeability tensors take the scalar values
 \begin{equation}
 \varepsilon_\rmr = \left\{ 
 \begin{array}{ll}
 \varepsilon_\rmc & \mbox{for} \, \bfx \in \Omega_\rmc  \\
 \varepsilon_\rmb & \mbox{for} \, \bfx \in \Omega_\rmb
 \end{array}\right. ,
 \quad
  \mu_\rmr = \left\{ 
 \begin{array}{ll}
 \mu_\rmc & \mbox{for} \, \bfx \in \Omega_\rmc \\
 \mu_\rmb & \mbox{for} \, \bfx \in \Omega_\rmb
 \end{array}\right.,
 \end{equation}
 where $\Omega_\rmc$ denotes the region inside the cylinder (of radius $r = a^\prime$) positioned in the centre of the fundamental unit cell $\Omega_\mathrm{WSC}$, and $\Omega_\rmb$ represents the remainder of the unit cell (i.e., $ \Omega_\rmb = \Omega_\mathrm{WSC} \backslash \Omega_\rmc$). Substituting the constitutive relations \eqref{seq:constit}  into the Maxwell system \eqref{seq:maxwelleqns} admits the vector wave equation
\begin{equation}
\label{seq:maxwelleig}
\nabla \times \varepsilon_\rmr^{-1} \nabla \times  \bfH -   \omegad^2 c_0^{-2} \mu_\rmr\bfH = 0.
\end{equation}
We then  impose that all electromagnetic fields have the dependence   $\mathrm{exp}(\rmi k_{z} z)$ and that  wave propagation in the direction of the fibres is forbidden, i.e.    $k_{z}=0$.  Subsequently,   the  Bloch vector takes the form $\bfk = (k_{x}, k_{y},0)$ where we introduce $\bfk_\rmB = (k_{x}, k_{y})$ as the in-plane Bloch wave vector.   This restriction on the wave vector  permits a decomposition of the    Maxwell wave equation \eqref{seq:maxwelleig} above  into two decoupled   scalar problems for the   two   orthogonal polarisation states; the first with the electric field $\bfE$ polarised in the direction $z$ (along the cylinders), and the second with the magnetic field $\bfH$ polarised in the direction $z$.  

  In this work, we examine the latter polarisation problem and subsequently    the Maxwell wave equation \eqref{seq:maxwelleig} reduces to the study of   the two-dimensional Helmholtz equation
 \begin{equation}
 \label{seq:helmholtz}
\nabla_\perp \cdot \left( \varepsilon_\rmr^{-1} \nabla_\perp H_\rmz \right) +  \omegad^2 c_0^{-2}  \mu_\rmr H_\rmz = 0,
 \end{equation}
 where $ \nabla_\perp \equiv(\partial_\rmx,\partial_\rmy)$ denotes the in-plane gradient, $\omegad$ denotes the angular   frequency, and $c_0$ is the speed of light in vacuum. Similarly,    the   electromagnetic continuity conditions at the edges of the cylinders   take the form
 \begin{subequations}
 \begin{equation}
 \label{seq:bcEM}
\left( H_\rmz^\rmc - H_\rmz^\rmb \right)\big|_{r=a^\prime} =0, \quad \left( \varepsilon_\rmc^{-1} \partial_r H_\rmz^\rmc - \varepsilon_\rmb^{-1} \partial_r H_\rmz^\rmb \right) \big|_{r =a^\prime} = 0, 
 \end{equation}
 corresponding to continuity of $H_z$ and $E_\theta$ at the   boundary, respectively.  Additionally, we   impose the Floquet--Bloch conditions  
 \begin{equation}
 H_\rmz(\bfx + \bfR_p)  = H_\rmz(\bfx) \rme^{\rmi \bfk_\rmB \cdot \bfR_p} ,
 \end{equation}
  \end{subequations}
 where $\bfR_p= (ma,na)$ for $m,n \in \mathbb{Z}$ is the real-lattice vector  for a square lattice of period $a$, where $p$ is multi-index notation for the pair $(m,n)$. The system for the   $E_\rmz$ polarised field is given by     exchanging $\varepsilon_\rmr$ and $\mu_\rmr$ in the above, and subsequently all band diagrams shown are thus also attainable under the $E_\rmz$ polarisation. The   general solution to the homogeneous two-dimensional Helmholtz equation    \eqref{seq:helmholtz} is well-known and takes the form 
  \begin{equation}
 H_\rmz = \left\{
\begin{array}{lc}
H_\rmz^\rmb  , & \mbox{for} \, \bfx \in \Omega_\rmb,\\
H_\rmz^\rmc  , & \mbox{for} \, \bfx \in \Omega_\rmc,
\end{array} 
  \right.
 \end{equation}
 where  
 \begin{subequations}
 \label{seq:homogcandb}
 \begin{align}
 H_\rmz^\rmb &= \sum\limits_{m=-\infty}^{\infty} \left[ \mathcal{A}_m J_m(  \omegand n_\rmb r) +  \mathcal{B}_m Y_m(\omegand n_\rmb r) \right]\rme^{\rmi m \theta},  \\
 H_\rmz^\rmc &= \sum\limits_{m=-\infty}^{\infty}  \mathcal{C}_m J_m(\omegand n_\rmc r) \, \rme^{\rmi m \theta}, 
 \end{align}
\end{subequations}
 with $(r, \theta)$ denoting polar coordinates in the $(x,y)$ plane, $\omegand = \omegad/c_0$, $n_\rmc = \sqrt{\varepsilon_\rmc} \sqrt{\mu_\rmc}$, and $n_\rmb = \sqrt{\varepsilon_\rmb} \sqrt{\mu_\rmb}$.  Note that in this work, we choose the positive branch of the square root function as the principal root and that $  \sqrt{\varepsilon} \sqrt{\mu} \neq \sqrt{\varepsilon \mu}$ in general; such a result only holds when both parameters are positive-valued. From our definition of $n$ above, we observe that both $\varepsilon$ and $\mu$ must be negative-valued in order to obtain a negative-index material ($n<0$).  The functions $J_m(z)$ and $Y_m(z)$ denote Bessel functions of the first and second kind. Substituting these general solutions  \eqref{seq:homogcandb} into the boundary conditions \eqref{seq:bcEM} we obtain
\begin{subequations}
\begin{equation}
 \mathcal{C}_m = \frac{ \mathcal{A}_m J_m(\omegand n_\rmb  a^\prime) +  \mathcal{B}_m Y_m(\omegand n_\rmb a^\prime) }{J_m(\omegand n_\rmc a^\prime) },
\end{equation}
and
\begin{equation}
\label{seq:AmMmBm}
 \mathcal{A}_m = -\scatcoeff_m \mathcal{B}_m,   
 \end{equation} 
where
  \begin{equation}
  \label{seq:M_m_polHz}
   \scatcoeff_m = \frac{    Z_\rmb^{-1} Y_m(n_\rmb \omegand a^\prime) \partial_r J _m(r) \big|_{r=n_\rmc \omegand a^\prime}   - Z_\rmc^{-1} J_m(n_\rmc \omegand a^\prime) \partial_r Y_m(r) \big|_{r=n_\rmb \omegand a^\prime}  } 
  {   Z_\rmb^{-1} J_m(n_\rmb \omegand a^\prime) \partial_r J_m (r) \big|_{r=n_\rmc \omegand a^\prime}    -Z_\rmc^{-1} J_m(n_\rmc \omegand a^\prime) \partial_r J_m (r) \big|_{r=n_\rmb \omegand a^\prime}  },
 \end{equation} 
\end{subequations}
having introduced  the impedances  $Z_\rmc = \sqrt{\mu_\rmc} / \sqrt{\varepsilon_\rmc}$ and $Z_\rmb = \sqrt{\mu_\rmb} / \sqrt{\varepsilon_\rmb}$.
In the presence of an incident plane wave, we determine the scattered fields  from all   other cylinders  in the array  (i.e., the multiple scattering contribution) using    Green's second identity,     admitting the  {\it Rayleigh identity}   \cite{movchan2002asymptotic}
  \begin{equation}
\label{seq:GADresult}
 \mathcal{A}_l =  \sum_{m=-\infty}^{\infty} (-1)^{l+m} S_{m-l}^\rmY \mathcal{B}_m,
 \end{equation}
where $S_m^\rmY$ denote lattice sums  (defined in Section \ref{sec:appendixA}) and we have omitted the incident field as we seek the Bloch modes of the structure.   Substituting \eqref{seq:AmMmBm} into \eqref{seq:GADresult} above we obtain the   system
 \begin{equation}
 \label{seq:dispeq}
 \scatcoeff_l  \mathcal{B}_l + \sum_{m=-\infty}^{\infty} (-1)^{l+m} S_{m-l}^\rmY  \mathcal{B}_m = 0,
 \end{equation}
 whose vanishing determinant   gives the dispersion equation for the homogeneous periodic problem. To obtain a numerically tractable form of \eqref{seq:dispeq},   it is necessary to   truncate parameters, i.e. to evaluate \eqref{seq:dispeq} for $l,m = -L, \ldots, L$.   For conventional photonic crystals, the convergence of the band surfaces   improves with larger truncation values $L$. However, at low frequencies    a monopole $(L=0)$ or dipole truncation ($L=1$) is     regarded as  a good approximation for the fully converged  dispersion equation.

 \section{Asymptotic analysis of dispersion equation} 
 In the preceding section we determined the form of the dispersion equation \eqref{seq:dispeq} for a two-dimensional photonic crystal of infinitely extending cylinders, which to monopolar order ($L=0$) takes the form
  \begin{equation}
 \label{seq:disprel_mono}
\scatcoeff_0 + S_0^\rmY  =0,
 \end{equation}
 and to dipolar order ($L=1$) takes the form
 \begin{equation}
 \label{seq:disprel}
 \left( \scatcoeff_0 + S_0^\rmY \right) \left( \left[ \scatcoeff_1 + S_0^\rmY \right]^2 - \left| S_2^\rmY \right|^2 \right) + \left( S_1^\rmY \right)^2  \left[ S_2^\rmY \right]^\ast  
- 2 \left| S_1^\rmY \right|^2 \left( \scatcoeff_1 + S_0^\rmY \right) 
+  \left( \left[ S_1^{\rmY } \right]^\ast \right)^2    S_2^\rmY =0,
 \end{equation}
where we have used the identities  $\scatcoeff_{-m} = \scatcoeff_m$  and $S_{-l}^\rmY = \left[ S_l^\rmY \right]^\ast $,  where $\ast$ denotes   complex conjugation  \cite{movchan2002asymptotic}.  In this section, we  obtain  conditions necessary for the existence of    band surfaces, through a close examination   of \eqref{seq:disprel_mono} and \eqref{seq:disprel}   in the low-frequency   regime.  
 	
\subsection{Asymptotic behaviour of  dispersion relations near the $\Gamma$ point}\label{sec:Gamma}
We begin by  evaluating asymptotic expressions for the scattering coefficients $N_m$ defined in  \eqref{seq:M_m_polHz} in the limit as $\omegand = \omegad/c_0 \rightarrow 0$.  These terms exhibit the leading order behaviour 
\begin{equation}
\label{seq:Mmlimit}
\lim_{\omegand\rightarrow 0}\scatcoeff_m = 
\left\{ \begin{array}{ll}
O(\omegand^{-2})\quad &\mbox{ for } m =0 \\
O( \omegand^{-2m})\quad &\mbox{ for } m >0  \\
 \end{array}\right. ,
\end{equation} 
and subsequently we  write  
\begin{subequations}
 \begin{align}
 \label{seq:Mexp}
 \lim_{{\omegand} \rightarrow 0} \left\{\scatcoeff_0 \right\}       \approx   \sum_{m=-2}^\infty D_m \omegand^m, \quad \mbox{ and } \quad
  \lim_{{\omegand} \rightarrow 0} \left\{ \scatcoeff_1 \right\}       \approx   \sum_{m=-2}^\infty E_m \omegand^m,
\end{align}
\end{subequations}
where   the first few terms of these series  are given by
\begin{subequations}
\begin{align}
\label{seq:M0asy}
D_{-2} &=    \frac{4}{\pi   a^{\prime 2} \varepsilon_\rmb} \frac{1}{\mu_\rmc - \mu_\rmb}    , \\  
\label{seq:M0asy2}
D_{0} &= \frac{    (4 \mu_\rmc  -3   \mu_\rmb)   \varepsilon_\rmb \mu_\rmb -\mu_\rmc^2  \varepsilon_\rmc }{2 \pi  \varepsilon_\rmb (\mu_\rmb-\mu_\rmc)^2} + \frac{2    }{  \pi     } \bigg[  \gamma_\rme +\log (\omegand)  \bigg] 
+ \frac{1}{  \pi   }   \log \left(\frac{a^{\prime 2} \varepsilon_\rmb \mu_\rmb }{4} \right) ,     \\
\label{seq:M1asy}
E_{-2} &=  \frac{4}{\pi  a^{\prime 2} \varepsilon_\rmb \mu_\rmb} \frac{\varepsilon_\rmc + \varepsilon_\rmb}{\varepsilon_\rmc - \varepsilon_\rmb}  , \\
\label{seq:M1asy2}
E_{0} &=   \frac{   \mu_\rmb (\varepsilon_\rmb + \varepsilon_\rmc)(5 \varepsilon_\rmc - 3 \varepsilon_\rmb)- 4  \mu_\rmc \varepsilon_\rmc^2}{2\pi \mu_\rmb (\varepsilon_\rmb - \varepsilon_\rmc)^2 }  
   + \frac{2}{ \pi} \bigg[  \gamma_\rme  + \log(\omegand) \bigg]  + \frac{1}{ \pi}  \log\left( \frac{ a^{\prime 2}  {\varepsilon_\rmb}   { \mu_\rmb}}{4} \right) , 
\end{align}
\end{subequations}
and $\gamma_\rme \approx 0.577215 \ldots$ denotes the Euler--Mascheroni constant. 
 
 Next we consider the   leading-order behaviours of the $S_l^\rmY$ terms present in \eqref{seq:disprel_mono} and \eqref{seq:disprel}. Semi-analytical representations for these sums were previously     determined by  \citet{mcphedran1996low,movchan2002asymptotic}, using  numerical estimates for certain terms. Using   the approach in \citet{chen2016evaluation} (details in Section \ref{sec:appendixA}) we obtain the fully closed forms 
\begin{subequations}
\label{seq:asySlYGamma}
\begin{align}
\lim_{k_\rmB \rightarrow 0}  \lim_{\omegand \rightarrow 0} S_0^Y &\approx - \frac{4}{a^2}\frac{1}{k_\rmB^2 - (n_\rmb\omegand)^2}  - \frac{2}{\pi} \log(n_\rmb\omegand )    
-   \frac{2 \gamma_\rme}{\pi}  - \frac{1}{ \pi} \log \left( \frac{ a^2 \pi}{ \Gamma(\tfrac{1}{4})^4 }\right)  ,  \\
\lim_{k_\rmB \rightarrow 0}  \lim_{\omegand \rightarrow 0} S_1^Y &\approx  - \frac{4 \rmi }{a^2}  \frac{k_\rmB}{n_\rmb\omegand} \frac{1}{k_\rmB^2 -     (n_\rmb\omegand)^2} \rme^{\rmi \theta_\rmB } + \frac{\rmi}{\pi} \frac{ k_\rmB}{n_\rmb\omegand  }   \rme^{\rmi \theta_\rmB }   ,  \\
\lim_{k_\rmB \rightarrow 0}  \lim_{\omegand \rightarrow 0} S_2^Y &\approx   \frac{4}{a^2} \frac{k_\rmB^2}{(n_\rmb\omegand)^2}     \frac{1}{k_\rmB^2 - (n_\rmb\omegand)^2} \rme^{2 \rmi \theta_\rmB}  
+  \frac{k_\rmB^2}{(n_\rmb\omegand)^2} \left[   \left( \frac{\Gamma(\tfrac{1}{4})^8 }{384 \pi^5} \right) \rme^{-2 \rmi \theta_\rmB}  - \frac{1}{2\pi}   \rme^{2 \rmi \theta_\rmB} \right],     
\end{align}
\end{subequations}
where $(k_\rmB,\theta_\rmB)$ is the polar representation of the Bloch vector $\bfk_\rmB$, and $\Gamma(z)$ is the Gamma function.  We now assume      that a band surface   emanates from the $\Gamma$ point at low frequencies in the form $\omegand = \alpha k_\rmB$, where $\alpha$ is real and positive-valued,   and substitute this    into  \eqref{seq:asySlYGamma} to obtain  
 \begin{align}
\label{seq:limSyNEW}
\lim_{k_\rmB \rightarrow 0}  \lim_{\omegand \rightarrow 0} S_l^Y(\omegand,\bfk_\rmB;n_\rmb) \sim O(\omegand^{-2}),
\end{align} 
for $l = 0,1,2,$ and where we treat $\log(\omegand)$ terms as $O(\omegand^0)$. Subsequently, we expand the   lattice sums in the form
  \begin{align}
 \label{seq:SlYexpG}
\lim_{k_\rmB \rightarrow 0} \lim_{\omegand \rightarrow 0} \left\{ S_0^Y \right\}   \approx   \sum_{m=-2}^\infty A_m \omegand^m , \qquad
\lim_{k_\rmB \rightarrow 0} \lim_{\omegand \rightarrow 0} \left\{ S_1^Y \right\}       \approx   \sum_{m=-2}^\infty B_m \omegand^m, \quad \mbox{ and } \quad
\lim_{k_\rmB \rightarrow 0} \lim_{\omegand \rightarrow 0} \left\{ S_2^Y \right\}       \approx   \sum_{m=-2}^\infty C_m \omegand^m,
\end{align}
 where summation is made   over non-zero orders, all of which are even due to symmetry. Substituting  \eqref{seq:Mexp} and \eqref{seq:SlYexpG}  into the monopolar dispersion equation \eqref{seq:disprel_mono} and collecting terms  in advancing orders of $\omegand$ reveals a hierarchy of equations describing the leading-order behaviour of the dispersion relation at low frequencies and the conditions necessary for the existence of the   band.

 The first non-vanishing order  for the monopolar dispersion equation \eqref{seq:disprel_mono} is given by
\begin{equation}
\label{seq:Okm2Gamma_coeffs}
O(\omegand^{-2}): \quad  A_{-2} + D_{-2} = 0,
\end{equation}
where \eqref{seq:Mexp} and \eqref{seq:asySlYGamma} admit  the dispersion relation  
\begin{equation}
\label{seq:disprelGamma_mono}
\omegand = \left\{ \frac{1}{ {   \varepsilon_\rmb \left[  \mu_\rmb + f   (\mu_\rmc - \mu_\rmb)  \right]} } \right\}^{1/2} k_\rmB,
\end{equation}
 Subsequently,   provided  $ \varepsilon_\rmb \mu_\rmb + f \varepsilon_\rmb (\mu_\rmc - \mu_\rmb)  >0$, a low-frequency band surface   emanates from the $\Gamma$ point within a monopolar approximation.

  Proceeding to the dipolar truncation \eqref{seq:disprel}, the first non-vanishing order is 
 \begin{multline}
\label{seq:Okm6Gamma}
O(\omegand^{-6}): \quad  (A_{-2})^3  + (A_{-2})^2 (D_{-2} + 2   E_{-2} ) + 
 A_{-2} \left[ (E_{-2})^2 - |C_{-2}|^2 + 2   D_{-2} E_{-2} -2 |B_{-2}|^2      \right] \\
+ (C_{-2})^\ast (B_{-2})^2 + \left[ (B_{-2})^\ast \right]^2 C_{-2}  
  - 2 |B_{-2}|^2 E_{-2}  + D_{-2} \left[ (E_{-2})^2 -|C_{-2}|^2 \right]= 0 ,
\end{multline}
which gives the low-frequency dispersion relation 
 
\begin{equation}
\label{seq:disprelGamma}
\omegand = \left\{ 
\frac{  { ( \varepsilon_\rmb + \varepsilon_\rmc)+f    ( \varepsilon_\rmb -   \varepsilon_\rmc ) }}{
 {  \left[\varepsilon_\rmb \mu_\rmb+f    \varepsilon_\rmb (\mu_\rmc-\mu_\rmb)\right]  } { \left[  (\varepsilon_\rmb+\varepsilon_\rmc)-f   (\varepsilon_\rmb-\varepsilon_\rmc)\right]  }} 
\right\}^{1/2} k_\rmB,
\end{equation}
 where $f = \pi a^{\prime 2}/a^2$ is the filling fraction, provided
\begin{equation}
\label{seq:Gammacondalpha}
\frac{  { ( \varepsilon_\rmb + \varepsilon_\rmc)+f    ( \varepsilon_\rmb -   \varepsilon_\rmc ) }}{
 {  \left[\varepsilon_\rmb \mu_\rmb+f    \varepsilon_\rmb (\mu_\rmc-\mu_\rmb)\right]  } { \left[  (\varepsilon_\rmb+\varepsilon_\rmc)-f   (\varepsilon_\rmb-\varepsilon_\rmc)\right]  }}>0.
\end{equation}
 Note that \eqref{seq:disprelGamma_mono} is recovered under the substitution   $\varepsilon_\rmc = \varepsilon_\rmb$ in \eqref{seq:disprelGamma}  above,  and that an effective refractive index is readily obtained via $n_\eff = 1/\alpha$. In this setting,  the expressions for the effective permittivity and permeability are  decoupled, motivating the representations
 \begin{equation}
 \varepsilon_\eff = \varepsilon_\rmb \left( \frac{(\varepsilon_\rmb+\varepsilon_\rmc)-f   (\varepsilon_\rmb-\varepsilon_\rmc)}{ ( \varepsilon_\rmb + \varepsilon_\rmc)+f    ( \varepsilon_\rmb -   \varepsilon_\rmc ) } \right)   = \varepsilon_\rmb + \frac{2\varepsilon_\rmb (\varepsilon_\rmc - \varepsilon_\rmb)f}{  (\varepsilon_\rmc + \varepsilon_\rmb) -    (\varepsilon_\rmc - \varepsilon_\rmb)f}, \qquad \mu_\eff =    \mu_\rmb+f      (\mu_\rmc-\mu_\rmb).
 \end{equation}
 Having determined the low-frequency dispersion relation   in the vicinity of the $\Gamma$ point, we now consider the asymptotics of the dispersion relation both at low frequencies and in the vicinity of other high symmetry points.

\subsection{Asymptotic behaviour of  dispersion relations near the $M$ point} \label{sec:SlYMpt}
The asymptotic forms of the first few  $S_l^\rmY$ sums   in the limit of vanishing frequency and  in the vicinity of the $M$ point     are given by
 \begin{subequations}
\label{seq:SmY}
\begin{align}
\label{seq:S0yM}
\lim_{k_\rmB \rightarrow M} \lim_{\omegand \rightarrow 0} \left\{ S_0^Y \right\}   &\approx    -\frac{2 \gamma_{\rme} }{\pi} - \frac{2}{\pi} \log\left(  n_\rmb\omegand  \right) 
- \frac{1}{\pi} \log\left( \frac{4 a^2 \pi}{\Gamma(\tfrac{1}{4})^4} \right),  \\
\label{seq:S1yM}
\lim_{k_\rmB \rightarrow M} \lim_{\omegand \rightarrow 0} \left\{ S_1^Y \right\}     & \approx \frac{\rmi}{\pi} \frac{ k_\rmB^\prime}{ n_\rmb\omegand } \rme^{\rmi \theta_\rmB^\prime}   , \\
\label{seq:S2yM}
\lim_{k_\rmB \rightarrow M} \lim_{\omegand \rightarrow 0} \left\{ S_2^Y \right\}     & \approx 
  -\frac{  k_\rmB^{\prime 2}  }{ ( n_\rmb\omegand)^2 }  \left[   \frac{1}{\pi^2}  \left( \frac{\Gamma(\tfrac{1}{4})^8}{128 \pi^3}    \right)  \rme^{-2 \rmi \theta_\rmB^\prime}
  +  \frac{1}{2\pi}        \rme^{2 \rmi \theta_\rmB^\prime} \right],  
\end{align}
\end{subequations}
 where $(k_\rmB^\prime ,\theta_\rmB^\prime )$ is the polar representation of $\bfk_\rmB^\prime = \bfk_\rmB - \bfM$, with $\bfM = (\pi/a,\pi/a)$.    A derivation of these expression  is presented in  Section \ref{sec:appendixA}.   As in  Section \ref{sec:Gamma} we now construct an ansatz assuming that a low frequency band surface emanates from the $M$ point in the form  $\omegand = \alpha^\prime k_\rmB^\prime$, where $\alpha^\prime$ is real and positive-valued. From the lattice sum asymptotics in \eqref{seq:SmY} this dependence yields
\begin{align}
\lim_{k_\rmB \rightarrow M}  \lim_{\omegand \rightarrow 0} S_l^Y(\omegand,\bfk_\rmB;n_\rmb) \sim   O(\omegand^{0}) ,  
 \end{align}
 and so   we introduce the expansions
 \begin{align}
 \label{seq:SlYexp}
\lim_{k_\rmB \rightarrow M} \lim_{\omegand \rightarrow 0} \left\{ S_0^Y \right\}  \approx   \sum_{m=0}^\infty A_m^\prime \omegand^m , \quad
\lim_{k_\rmB \rightarrow M} \lim_{\omegand \rightarrow 0} \left\{ S_1^Y \right\}      \approx   \sum_{m=0}^\infty B_m^\prime  \omegand^m,\quad \mbox{ and } \quad
\lim_{k_\rmB \rightarrow M} \lim_{\omegand \rightarrow 0} \left\{ S_2^Y \right\}       \approx   \sum_{m=0}^\infty C_m^\prime  \omegand^m,
\end{align}
where summation is made over even orders, as the odd orders remain identically zero.  Next we substitute the expansion for $\scatcoeff_{0}$ in \eqref{seq:Mexp} and $S_{0}^\rmY$ in  \eqref{seq:SlYexp}   into the monopolar dispersion equation \eqref{seq:disprel_mono} and collect terms  in $\omegand$ to obtain a hierarchy of conditions. The first two of these equations are  
\begin{subequations}
\label{seq:hierarchyMmono}
\begin{align}
O(\omegand^{-2}): &\quad D_{-2} = 0, \\
O(\omegand^{0}): &\quad D_{0} + A_0^\prime  = 0,
\end{align}
\end{subequations}
where from \eqref{seq:M0asy}, it follows that the $O(\omegand^{-2})$ equation is never satisfied for finite permittivity and permeability values, and subsequently a band surface is not supported from the $M$ point at low frequencies within a monopolar approximation. This finding is consistent with the fact that a monopole field is   symmetric and subsequently  cannot satisfy quasi-periodic boundary conditions at the unit cell edges.

In contrast, from  the dipolar dispersion equation \eqref{seq:disprel} we   obtain  
\begin{subequations}
\label{seq:MOkeqns}
\begin{align}
\label{seq:Okm6}
O(\omegand^{-6}): &\quad \left(   E_{-2} \right)^2  D_{-2} = 0, \\
\label{seq:Okm4}
O(\omegand^{-4}): &\quad  E_{-2} \left[  (A_0^\prime  + D_0) E_{-2}   + 2  (A_0^\prime  + E_0 )  D_{-2}  \right] = 0,   \\  \label{seq:Okm2}  
O(\omegand^{-2}):   &\quad \begin{multlined}[t]  D_{-2} \left[ (A_0^\prime  + E_0) ^2  -|C_0^\prime |^2    \right]  
+  (E_{-2})^2 (A_2^\prime  + D_2)  \\
  + 2 E_{-2} \left[   (A_0^\prime)^2     -  |B_0^\prime |^2  + A_2^\prime  D_{-2} +D_0   E_0 + A_0^\prime  ( D_0  +       E_0) + D_{-2}  E_2 \right] =0, \end{multlined}
\end{align}
\end{subequations}
 where from inspection,   the $O(\omegand^{-6})$ and $O(\omegand^{-4})$ equations are     satisfied when $E_{-2} = 0$, or  $\varepsilon_\rmc = -\varepsilon_\rmb$ following \eqref{seq:M1asy}.
 Furthermore, the $O(\omegand^{-2})$ equation is simplified by  the  $E_{-2} = 0$ requirement, reducing it to the form 
\begin{align}
\label{seq:Okm2redM}
O(\omegand^{-2}):  \quad D_{-2} \left(  A_0^\prime  + E_0   - |C_0^\prime | \right)  \left(  A_0^\prime + E_0   + |C_0^\prime | \right) = 0,
\end{align}
where   $A_0^\prime  + E_0   + |C_0^\prime| = 0$ ultimately gives the dispersion relation 
 
 \begin{equation}
 \label{seq:kvk0}
   \omegand  = \left\{ \frac{1}{  8 \pi^{ 2}  } \left| 64 \pi^4 \rme^{4 \rmi \theta_\rmB^\prime} + \Gamma(\tfrac{1}{4} )^8 \right|^{1/2}  \left[ \varepsilon_\rmb \mu_\rmc + 2 \varepsilon_\rmb \mu_\rmb
        \log\left( \frac{16   \pi^2}{f   \Gamma(\tfrac{1}{4})^4}  \right) \right]^{-1/2}  \right\} k_\rmB^{\prime  }.
\end{equation}
  Thus within a dipolar approximation, a band surface is supported from the $M$ point at   low frequencies (with a slope described by \eqref{seq:kvk0}) provided  both
  \begin{subequations}
  \begin{equation}
  \label{seq:epsccond0M} 
  \varepsilon_\rmc = -\varepsilon_\rmb, 
  \end{equation}
   and   
\begin{equation}
\label{seq:epsccond1}
 \mu_\rmc 
        >\left[ \log\left( \frac{f^2  \Gamma(\tfrac{1}{4})^8}{2^8   \pi^4}  \right) \right]  \mu_\rmb , 
     \end{equation} 
       \end{subequations}
  are satisfied, where the prefactor in \eqref{seq:epsccond1} is negative-valued for $0<f\lesssim0.914$ (and for reference, $f = \pi/4 \approx 0.785$ represents   dense packing   for a square array of cylinders).  The expression  \eqref{seq:epsccond1} follows from the condition that  $\alpha$ must be real and positive-valued.

 \subsection{Asymptotic behaviour of  dispersion relations near the $X$ point} \label{sec:SlYXpt}
The asymptotic  forms of the first few    $S_l^\rmY$at both low frequencies and  in   the vicinity of the $X$ point   are given by
\begin{subequations}
\label{seq:asySlYX}
\begin{align}
  \lim_{k_\rmB \rightarrow X}  \lim_{\omegand \rightarrow 0} S_0^Y  &\approx   -   \frac{2 \gamma_\rme}{\pi}  - \frac{2}{\pi} \log(n_\rmb\omegand ) 
- \frac{1}{ \pi} \log \left( \frac{ 8 a^2 \pi}{ \Gamma(\tfrac{1}{4})^4 }\right) ,  \\ 
   \lim_{k_\rmB \rightarrow X}  \lim_{\omegand \rightarrow 0} S_1^Y  &\approx 
  \frac{\rmi}{ \pi}   \frac{ k_\rmB^\pp}{n_\rmb\omegand} \left(  \rme^{\rmi \theta_\rmB^\pp}  +  \frac{     \Gamma  (\tfrac{1}{4} )^4 }{8 \pi^2}  \rme^{-\rmi \theta_\rmB^\pp} \right) , \\
  \lim_{k_\rmB \rightarrow X}  \lim_{\omegand \rightarrow 0} S_2^Y  &\approx 
  \frac{1}{ (n_\rmb\omegand)^2} \left( \frac{ \Gamma(\tfrac{1}{4})^4}{4 a^2 \pi^2} \right) 
  + \frac{\Gamma(\tfrac{1}{4})^4 }{8 \pi^3}\left(1- \frac{k_\rmB^{\pp 2} }{  (n_\rmb\omegand)^2 }\right)  
  +   \frac{k_\rmB^{\pp 2}}{ (n_\rmb\omegand)^2}  \left[ \frac{1}{\pi^2} \left( \frac{  \Gamma(\tfrac{1}{4})^8 }{128  \pi^3}  \right)\rme^{-2 \rmi \theta_\rmB^\pp} 
  -\frac{ 1  }{2 \pi} \rme^{2 \rmi \theta_\rmB^\pp} 
    \right]  ,  
\end{align}
\end{subequations}
 where $(k_\rmB^{\pp},\theta_\rmB^\pp)$ is the polar representation of $\bfk_\rmB^\pp = \bfk_\rmB - \bfX$. The derivation for the   lattice sum  expressions above is given in  Section \ref{sec:appendixA}.  In an identical manner to Sections \ref{sec:Gamma} and \ref{sec:SlYMpt},  we  now assume that a linear band surface emanates from the $X$ point at low frequencies. That is, we  substitute the form $\omegand = \alpha^{\prime \prime} k_\rmB^\pp$, where $\alpha^{\prime \prime}$ is real and positive-valued,   into \eqref{seq:asySlYX} and observe that 
 \begin{align*}
 \lim_{k_\rmB \rightarrow X}  \lim_{\omegand \rightarrow 0} S_l^Y(\omegand,\bfk_\rmB;n_\rmb) &\sim O(\omegand^{-\mathrm{floor}(l^2/2)}),
 \end{align*}
for $l=0,1,2$.   Such limit behaviour     admits  the   expansions
\begin{subequations}
  \begin{align}
 \label{seq:SlYexpX}
\lim_{k_\rmB \rightarrow X} \lim_{\omegand \rightarrow 0} \left\{ S_0^Y \right\}   \approx   \sum_{m=0}^\infty A_m^{\prime\prime} \omegand^m , \quad
\lim_{k_\rmB \rightarrow X} \lim_{\omegand \rightarrow 0} \left\{ S_1^Y \right\}       \approx   \sum_{m=0}^\infty B_m^{\prime\prime} \omegand^m,\quad \mbox{ and }\quad 
\lim_{k_\rmB \rightarrow X} \lim_{\omegand \rightarrow 0} \left\{ S_2^Y \right\}       \approx   \sum_{m=-2}^\infty C_m^{\prime\prime} \omegand^m,
\end{align}
\end{subequations}
where summation is made over even orders, as the odd orders remain identically zero. For the monopolar dispersion equation,  we   obtain an analogous hierarchy of equations to \eqref{seq:hierarchyMmono} which does not permit a low frequency band at $X$. However, for the dipolar dispersion relation we obtain  
\begin{subequations}
\begin{align}
\label{seq:Okm6X}
O(\omegand^{-6}): &\;  D_{-2} \left( E_{-2} - |C_{-2}^{\prime\prime}| \right)  \left( E_{-2}+ |C_{-2}^{\prime\prime}| \right) = 0, \\
\label{seq:Okm4X}
O(\omegand^{-4}): &\;     |C_{-2}^{\prime\prime}|^2 (A_0^{\prime\prime} + D_0) + 2 |C_{-2}^{\prime\prime}| D_{-2} |C_0^{\prime\prime} |   - E_{-2} (D_0 E_{-2} + A_0^{\prime\prime} (2 D_{-2} + E_{-2})  
+ 2 D_{-2} E_0) = 0,  
\end{align}
\end{subequations}
where the $O(\omegand^{-6})$ system is satisfied provided  $ |C_{-2}^{\prime\prime}|=  -E_{-2} $ or equivalently
\begin{equation}
\label{seq:condCm2Em2}
\varepsilon_\rmc = \left( \frac{ f   \Gamma(\tfrac{1}{4})^4 - 16   \pi^2 }{ f   \Gamma(\tfrac{1}{4})^4+ 16  \pi^2} \right) \varepsilon_\rmb.
\end{equation}
The above condition   simplifies  the $O(\omegand^{-4})$ expression \eqref{seq:Okm4X} to the form
\begin{equation}
\label{seq:Okm4x}
 2 D_{-2} E_{-2} \left(   A_0^{\prime\prime} + E_0 +  |C_0^{\prime\prime}| \right) =0.
\end{equation}
   Substituting \eqref{seq:M1asy2}, the coefficients for \eqref{seq:asySlYX}, and \eqref{seq:condCm2Em2} into the reduced condition $  A_0^{\prime\prime} + E_0 +  |C_0^{\prime\prime}|  =0$      yields  a lengthy expression of the form
\begin{equation}
\label{seq:simpl}
\beta_1  + \beta_2   \left| \frac{\beta_3 + \beta_4  \alpha^{\pp 2}}{\alpha^{\pp 2}} \right| = 0,
\end{equation}
where $\beta_j$ are constants.   To ensure that   $\beta_1 = 0$ it is necessary that
\begin{equation}
\label{seq:condepscX}
\mu_\rmc = \left( \frac{   f   \Gamma(\tfrac{1}{4})^4 ( f    \Gamma(\tfrac{1}{4})^4  - 64     \pi^2 )  
 + 512   \pi^4 \log\left( {f   \Gamma(\tfrac{1}{4})^4}/{(32   \pi^2) }\right)   }{(f   \Gamma(\tfrac{1}{4})^4  - 16 \pi^2)^2}  \right)\mu_\rmb,
\end{equation}
which then simplifies  \eqref{seq:simpl} to reveal  the low-frequency dispersion relation
\begin{subequations}
\begin{equation}
\label{seq:kk0X}
\omegand = \left(  \frac{ \left( 16 \Gamma(\tfrac{1}{4})^4 \pi^2+ 64 \pi^4 \rme^{2\rmi \theta_\rmB^{\pp}} -\Gamma(\tfrac{1}{4})^8 \rme^{-2\rmi \theta_\rmB^{\pp}}   \right) }{16 \Gamma(\tfrac{1}{4})^4 \pi^2 \varepsilon_\rmb \mu_\rmb} \right)^{1/2} k_\rmB^\pp.  
\end{equation}
However, we remark that the $\alpha^{\prime\prime}$  in \eqref{seq:kk0X} is only real-valued  along the paths $\Gamma X$ and $XM$ (i.e., for $\theta_\rmB^\pp = \pi$ and $\theta_\rmB^\pp =\pi/2$, respectively). Numerical investigations confirm the existence of a band surface across the entire Brillouin zone,  where the contours are elliptical at low frequencies. After amending our dispersion relation ansatz to   $\omegand^2 = \alpha_\rmx (k_{\rmB \rmx}-\pi/a)^2 + \alpha_\rmy k_{\rmB \rmy}^2$ and using \eqref{seq:kk0X} above we finally obtain  
 
\begin{equation}
\label{seq:disprelXellipse}	 
\omegand^2 = \left( \frac{ 16 \Gamma(\tfrac{1}{4})^4 \pi^2 + 64 \pi^4 - \Gamma(\tfrac{1}{4})^8 }{16 \Gamma(\tfrac{1}{4})^4  \pi^2 \varepsilon_\rmb \mu_\rmb} \right) k_{\rmB \rmx}^{\pp 2} + \left( \frac{ 16 \Gamma(\tfrac{1}{4})^4 \pi^2 - 64 \pi^4 + \Gamma(\tfrac{1}{4})^8 }{16 \Gamma(\tfrac{1}{4})^4  \pi^2 \varepsilon_\rmb \mu_\rmb} \right)  k_{\rmB \rmy}^{\pp 2},
\end{equation}
 for the first band surface as $\omegand \rightarrow 0$ and  for $\bfk_\rmB$ in the vicinity of $X$. Subsequently, the requirement that $\alpha_{\rmx},\alpha_\rmy>0$ is equivalent to $\varepsilon_\rmb \mu_\rmb >0$.
\end{subequations}

\section{Low-frequency descriptions for non-magnetic media} \label{sec:nonmag}

In this section, we provide a condensed outline of the conditions for a band to emanate from the $\Gamma$, $X$, and $M$ points of the crystal  when it comprises non-magnetic media, within a dipolar approximation.   As mentioned in Section \ref{sec:problemform} the leading-order   behaviour of the scattering coefficients $\scatcoeff_m$ differ  for photonic crystals made from non-magnetic and magnetic materials. In the former instance, the scattering coefficients take the form
 \begin{equation}
\lim_{\omegand\rightarrow 0}\scatcoeff_m = 
\left\{ \begin{array}{ll}
O(\omegand^{-4})\quad &\mbox{ for } m =0 \\
O(\omegand^{-2m})\quad &\mbox{ for } m >0  \\
 \end{array}\right. ,
\end{equation} 
in contrast to \eqref{seq:Mmlimit}. Expanding the first two  $\scatcoeff_n$ coefficients    as
 \begin{align}
 \label{seq:Mexpnonmag}
 \lim_{\omegand \rightarrow 0} \left\{\scatcoeff_0 \right\}       \approx   \sum_{m=-4}^\infty \widetilde{D}_m \omegand^m,\quad \mbox{ and } \quad
  \lim_{\omegand \rightarrow 0} \left\{ \scatcoeff_1 \right\}       \approx   \sum_{m=-2}^\infty \widetilde{E}_m \omegand^m,
\end{align}
we obtain the first few terms 
\begin{subequations}
\label{seq:Mmnonmag}
\begin{align}
  \widetilde{D}_{-4}    &=    \frac{32}{\pi   a^{\prime 4} \varepsilon_\rmb} \frac{1}{\varepsilon_\rmc - \varepsilon_\rmb}    , \\
 \widetilde{D}_{-2}  &= \frac{16 }{3 \pi a^{\prime 2} \varepsilon_\rmb } \frac{ \varepsilon_\rmc - 2 \varepsilon_\rmb}{\varepsilon_\rmb - \varepsilon_\rmc},     \\
 \widetilde{E}_{-2}    &=  \frac{4}{\pi  a^{\prime 2} \varepsilon_\rmb  } \frac{\varepsilon_\rmc + \varepsilon_\rmb}{\varepsilon_\rmc - \varepsilon_\rmb}  , \\
 \widetilde{E}_{0}   &=  \frac{1}{2\pi } \left[ \frac{ \varepsilon_\rmc + 3 \varepsilon_\rmb  }{ \varepsilon_\rmc - \varepsilon_\rmb } + 4  \gamma_\rme      +2   \log\left( \frac{a^{\prime 2}   \varepsilon_\rmb}{4} \right) + 4  \log(\omegand) \right].  
\end{align}
\end{subequations}
Substituting the $\scatcoeff_m$ expansions \eqref{seq:Mexpnonmag}   and  the  $\Gamma$ centred   lattice sum expansions \eqref{seq:asySlYGamma} into the dipolar dispersion relation \eqref{seq:disprel} gives  an $O(\omegand^{-8})$ condition, and not an $O(\omegand^{-6})$ condition as before, revealing
\begin{equation}
\label{seq:Gcondnonmag}
\omegand = \left\{ \frac{1}{\varepsilon_\rmb}\frac{(\varepsilon_\rmb + \varepsilon_\rmc) + f (\varepsilon_\rmb - \varepsilon_\rmc)}{(\varepsilon_\rmb + \varepsilon_\rmc) - f (\varepsilon_\rmb - \varepsilon_\rmc)}\right\}^{1/2} k_\rmB,
\end{equation}
which is identical to  \eqref{seq:disprelGamma} above with the replacement $\mu_\rmb, \mu_\rmc \mapsto 1$.

 Similarly, for the $M$ point, substituting \eqref{seq:Mexpnonmag} and \eqref{seq:SmY} into the dipolar dispersion equation we obtain a system of conditions analogous to \eqref{seq:MOkeqns}. For non-magnetic crystals, the $O(\omegand^{-8})$ and $O(\omegand^{-6})$ conditions are   satisfied when 
 \begin{subequations}
\begin{equation}
\label{seq:Mcondnonmagz}
\varepsilon_\rmc = -\varepsilon_\rmb,
\end{equation}
 with the $O(\omegand^{-4})$ condition giving
\begin{equation}
\label{seq:Mcondnonmag}
   \omegand  = \left\{ \frac{1}{  8 \pi^{ 2}  } \left| 64 \pi^4 \rme^{4 \rmi \theta_\rmB^\prime} + \Gamma(\tfrac{1}{4} )^8 \right|^{1/2}  \left[ \varepsilon_\rmb   + 2 \varepsilon_\rmb 
        \log\left( \frac{16   \pi^2}{f   \Gamma(\tfrac{1}{4})^4}  \right) \right]^{-1/2}  \right\} k_\rmB^{\prime  },
\end{equation}
 \end{subequations}
which is identical to \eqref{seq:kvk0} but with $\mu_\rmb,\mu_\rmc \mapsto 1$. The analogue to $\alpha>0$ in \eqref{seq:epsccond1} now takes the simpler form  $\varepsilon_\rmb  >0$.  For the $X$ point,  the leading-order $O(\omegand^{-8})$ condition (c.f.  $O(\omegand^{-6})$ before) is satisfied for  
 \begin{subequations}
\begin{equation}
\label{seq:xcondnonmag}
\varepsilon_\rmc = \left( \frac{ f     \Gamma(\tfrac{1}{4})^4 - 16   \pi^2 }{ f       \Gamma(\tfrac{1}{4})^4 + 16  \pi^2} \right) \varepsilon_\rmb,  
\end{equation}
(which is identical to \eqref{seq:condCm2Em2}). However, the $O(\omegand^{-6})$ condition  (c.f.  $O(\omegand^{-4})$ before) takes the form  \eqref{seq:simpl} where $\beta_1 = 0$ requires that 
\begin{equation}
\label{seq:conditionXnonmagnotsupp}
 f  \Gamma(\tfrac{1}{4})^4   + 8\pi^2 + 16\pi^2\log \left( \frac{32 \pi^2}{f   \Gamma(\tfrac{1}{4})^4}\right) = 0,
\end{equation}
 \end{subequations}
which does not hold for any $f$. Hence, non-magnetic photonic crystals do not support first bands emerging from the $X$ point as  $\omegand = \alpha^{\prime \prime} k_\rmB^\pp$ or $\omegand^2 = \alpha_\rmx (k_{\rmB \rmx}-\pi/a)^2 + \alpha_\rmy k_{\rmB \rmy}^2$. To clarify,  \eqref{seq:conditionXnonmagnotsupp} demonstrates that the ansatz $\omegand = \alpha^{\prime \prime} k_\rmB^\pp$ in a non-magnetic crystal cannot satisfy $\beta_1 = 0$ in one of the orders, and subsequently, a dispersion relation cannot be extracted. Hence, that particular ansatz does not describe a low-frequency behaviour of non-magnetic crystals. From our numerical study, we have observed     $\omegand = C_1 \, |\bfk_\rmB| + C_2 \,|\bfk_\rmB - \bfX|$ in non-magnetic crystals,  which   is an entirely different ansatz, with different low-frequency behaviour, that will undoubtedly be supported. However, we believe that obtaining    descriptions for the latter case   falls outside the scope of the present work. Note that for non-magnetic crystals, if we have negative permittivities in the constituent materials, then the refractive index and impedance are purely imaginary.

\section{Numerical results} \label{sec:numericalsec}
In this section, we provide further details on the numerical calculation of the  band structures     of a two-dimensional  photonic crystal. For this task, we     search for the vanishing determinant of the system \cite{movchan2002asymptotic,poulton2000eigenvalue}
      \begin{equation}
\label{seq:antipov}
      \left( \delta_{lm} +  \frac{\mathrm{sgn}(\scatcoeff_l) }{ (|\scatcoeff_l| |\scatcoeff_m|)^{1/2}} (-1)^{l+m} S_{m-l}^\rmY \right) \mathcal{D}_m = 0,
      \end{equation}
            where $\delta_{lm}$ denotes the Kronecker delta function and  $\mathcal{D}_m = (  |\scatcoeff_m|)^{1/2} \mathcal{B}_m$,  in place of   the original system \eqref{seq:dispeq}.  The above representation scales the value of the determinant in absolute value terms, as the scattering coefficients and lattice sum terms   become increasingly large for $l,m \rightarrow \infty$. The lattice sums $S_l^\rmY$ are evaluated using  the expressions    given in  \citet{mcphedran2000lattice}, which  explicitly relate  one-    and two-dimensional array sums for any Bravais lattice configuration. For the one-dimensional array sums,  we use   \citet[(2.53) and (2.54)]{linton1998greens}, which are   accelerated forms of the   expressions first derived in \citet{twersky1961elementary}, and then evaluate the necessary correction factors in \citet{mcphedran2000lattice} to obtain two-dimensional sums for a square lattice. The approach of evaluating grating sums and then correction factors  \cite{mcphedran2000lattice} is preferred to evaluating     accelerated lattice sum expressions   \cite{chin1994greens,mcphedran1996low,movchan2002asymptotic} directly, as we have found   that the former   approach   is much more numerically stable    for $\omegand\rightarrow 0$.

  \begin{figure}[t]
 
     \subfloat[\label{fig:modecomparisonGXM1}]{%
       \includegraphics[width=0.325\textwidth]{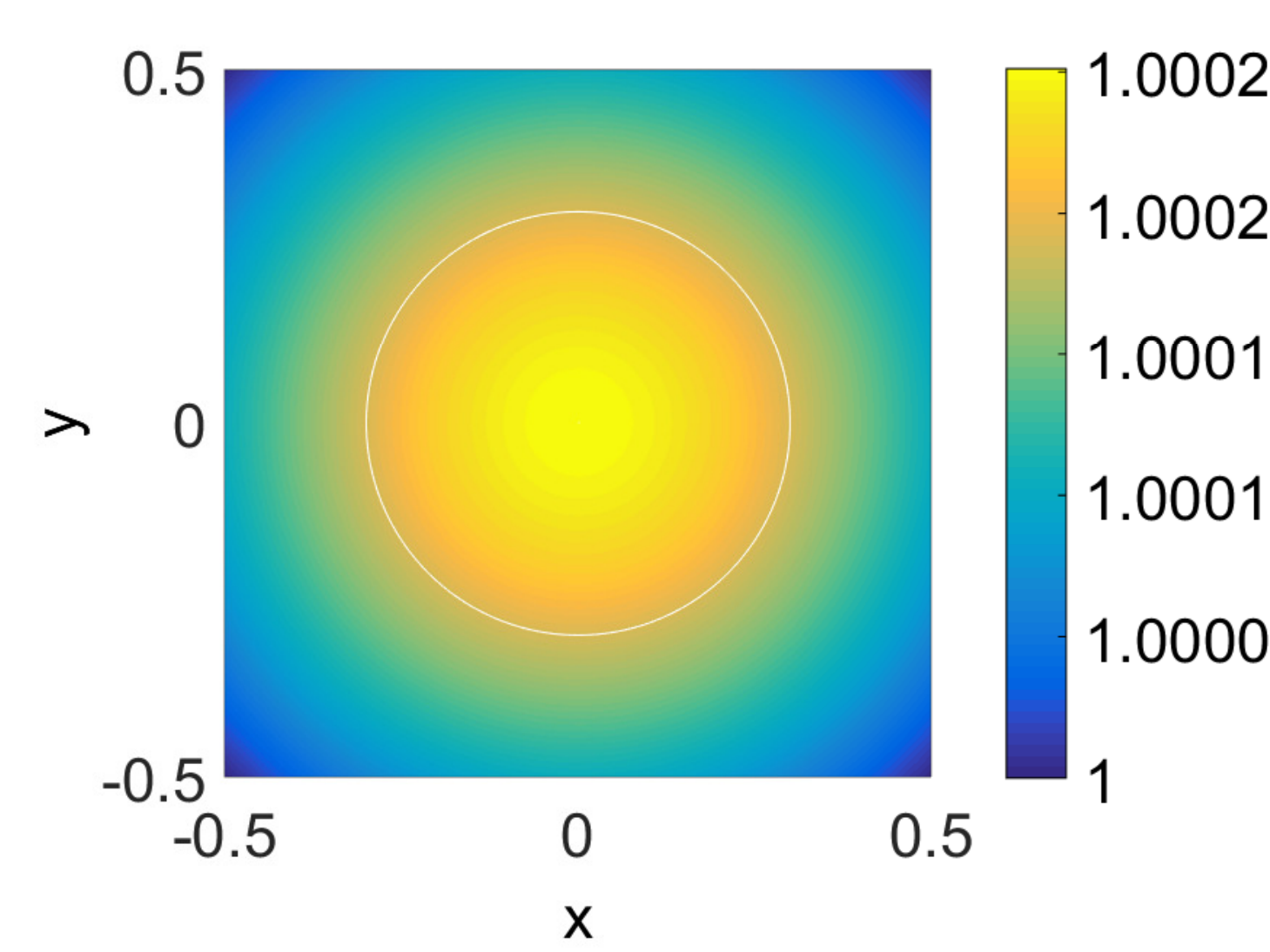}
     }
     \hfill
     \subfloat[\label{fig:modecomparisonGXM2}]{%
       \includegraphics[width=0.325\textwidth]{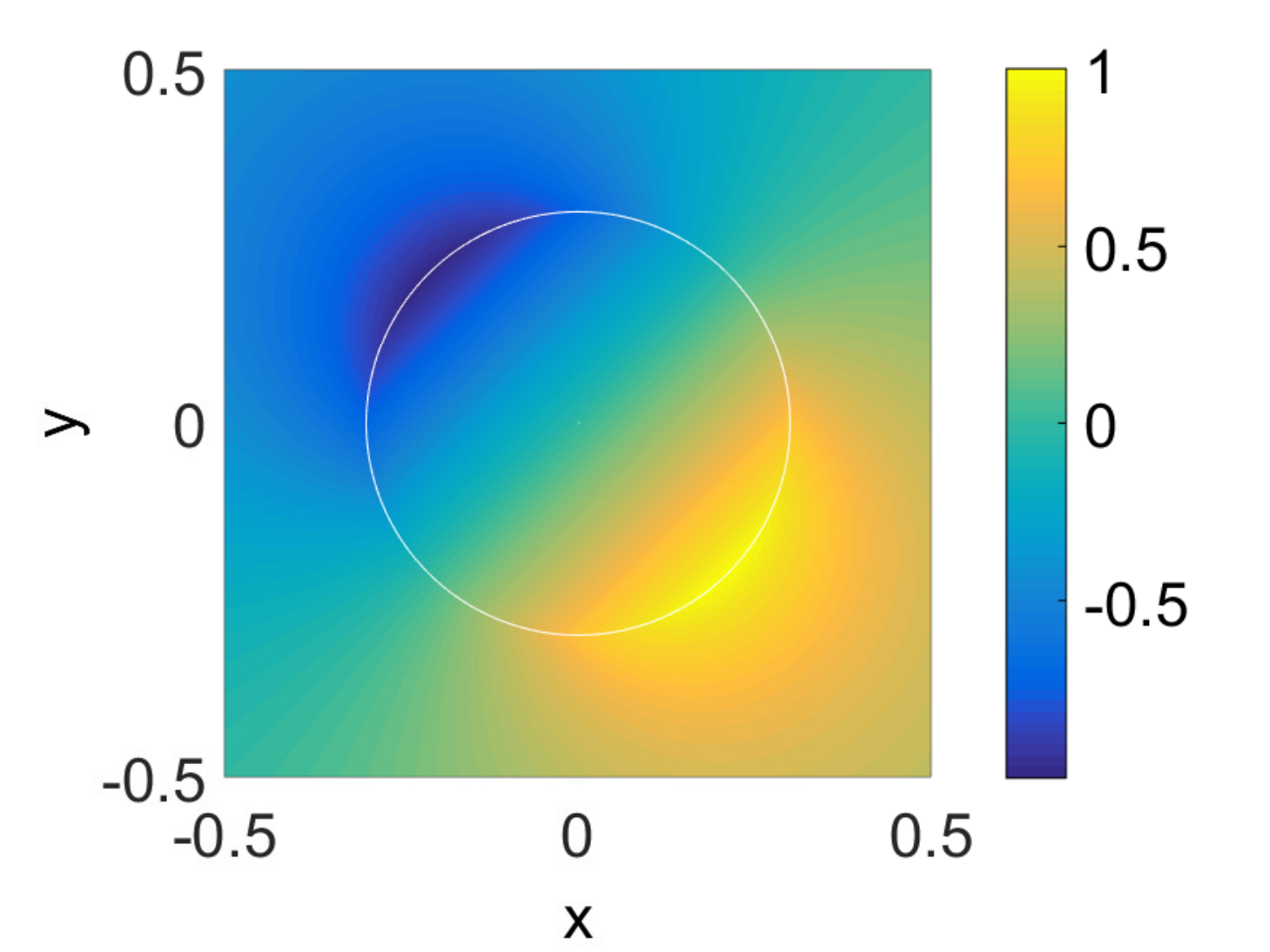}
     }
       \hfill
     \subfloat[\label{fig:modecomparisonGXM3}]{%
       \includegraphics[width=0.325\textwidth]{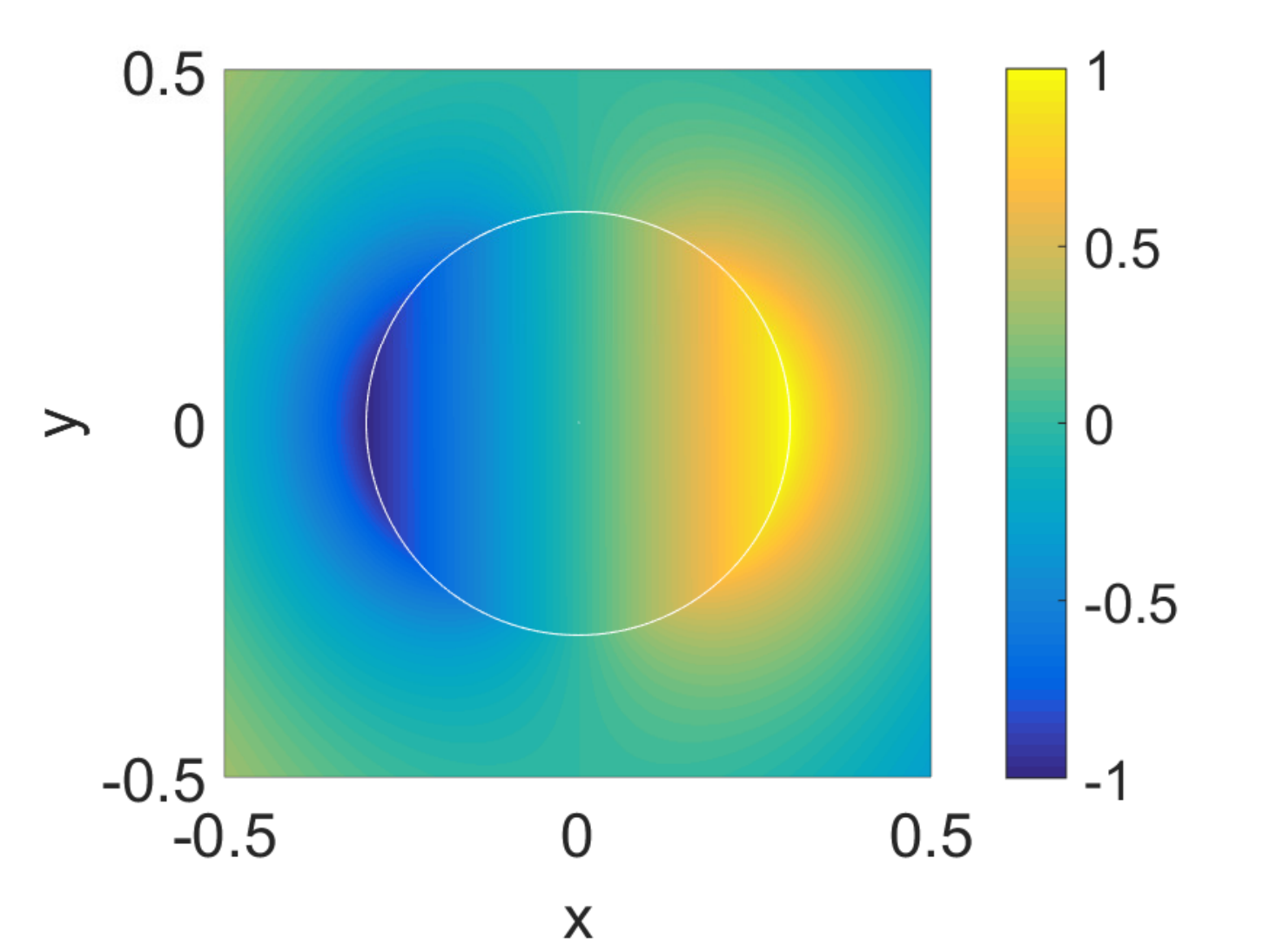}
     }
     
         \caption{ Representative modes near the origins of the \protect\subref{fig:modecomparisonGXM1} $\Gamma$, \protect\subref{fig:modecomparisonGXM2} $M$, and  \protect\subref{fig:modecomparisonGXM3} $X$ emanating band surfaces (for $\varepsilon_\rmc =  1$ and $\mu_\rmc = 2$,  $\varepsilon_\rmc = -1$ and $\mu_\rmc = 2$, and $\varepsilon_\rmc \approx -0.53$ and $\mu_\rmc \approx -10.22$ embedded in air, respectively).  In all instances, the normalised $\mathrm{Re}(H_\rmz)$ component of the field is given for lattice period    $a=1$,   cylinder radius   $a^\prime = 0.3a$ (boundary marked in white), and $L=1$. \label{fig:modecomparisonGXM}   }
   \end{figure}

 For reference, in Fig.~\ref{fig:modecomparisonGXM} we demonstrate the shape of the modes in the vicinity of their emanation points. In Fig.~\ref{fig:modecomparisonGXM1} we present the mode for $\varepsilon_\rmc = 1$ and $\mu_\rmc = 2$ in air with $\bfk_\rmB \approx(0,0)$ which shows that the mode is approximately constant as $\omegand\rightarrow 0$. This behaviour is typical  for a crystal with such a low material contrast \cite{bensoussan1978asymptotic,jikov2012homogenization}. In Fig.~\ref{fig:modecomparisonGXM2}, we have $\varepsilon_\rmc = -1$ and $\mu_\rmc = 2$ in air with $\bfk_\rmB \approx(\pi/a,\pi/a)$ which shows that the mode takes the form of a line dipole largely concentrated to the boundary of the cylinder. The dipole response is  also oriented along $\theta = 3\pi/4$ to ensure that    the   Bloch  conditions are met  along all edges of the unit cell (i.e., anti-periodicity). Finally, in Fig.~\ref{fig:modecomparisonGXM3} we give the mode for $\varepsilon_\rmc \approx -0.53$ and $\mu_\rmc \approx -10.22$ in air with $\bfk_\rmB \approx(\pi/a,0)$ which also shows a    dipole response concentrated to the boundary of the cylinder. This dipole response      is oriented along $\theta = \pi/2$ which ensures that  the Bloch conditions are satisfied.

In Fig.~\ref{fig:comparebands} we present  band diagrams for a crystal satisfying the conditions for $\Gamma$, $M$, and $X$ point emanation (given by \eqref{seq:Gammacondalpha},   \eqref{seq:epsccond0M} and \eqref{seq:epsccond1}, and \eqref{seq:condCm2Em2} and \eqref{seq:condepscX}, respectively), but with a higher truncation  value $L=5$. In all instances, the dashed red line represents the dipolar estimate of the first band surface. In Fig.~\ref{fig:comparebands1}  we observe that there   are negligible differences between the few first band surfaces of a conventional photonic crystal, upon comparing  Fig.~\ref{fig:comparebands1} ($L=5$) with Fig.~1a of the letter ($L=1$).  In Fig.~\ref{fig:comparebands2},   we impose the $M$-point conditions for $L=1$ and evaluate the dispersion equation for $L=5$, and see   that  the   conditions correctly predict the material parameters for which $M$ emanation occurs, but that the asymptotics are unable to correctly predict the slope.  The higher bands are also different, but this is not unexpected as  higher order multipole terms are generally required to reproduce both the dispersion relation and the modal fields at higher frequencies.  We emphasize that this   sensitivity is observed close to the  $M$ point conditions {\it alone} (and is not observed      away from $\varepsilon_\rmc =-\varepsilon_\rmb$), due to its correspondence with the anomalous resonance condition  \cite{bergman1979dielectric,mcphedran1980electrostatic}.  We discuss this in  further detail below. In Fig.~\ref{fig:comparebands3}  we impose  the $X$-point conditions corresponding to $L=1$ for   $L=5$, where we observe that the first two bands are well-approximated, but that the origin of the first band moves slightly away from $X$. Note that the  first band is   re-centred about $X$ after slightly  perturbing the values of $\varepsilon_\rmc$ and  $\mu_\rmc$. These variations in the band diagrams   emphasise  that  the conditions and descriptions we obtain for $\Gamma$  and $X$ point emanation are good approximations for the full systems, but that our description of the first band emerging {\it from $M$} are {\it   specific to the dipolar    dispersion equation} and are {\it not necessarily  accurate} for higher truncations.      Analytically  determining high-truncation conditions   poses a significant  challenge, as  the   dispersion equation becomes highly intractable   for large $L$.

 \begin{figure}[t]
 
     \subfloat[\label{fig:comparebands1}]{%
       \includegraphics[width=0.3225\textwidth]{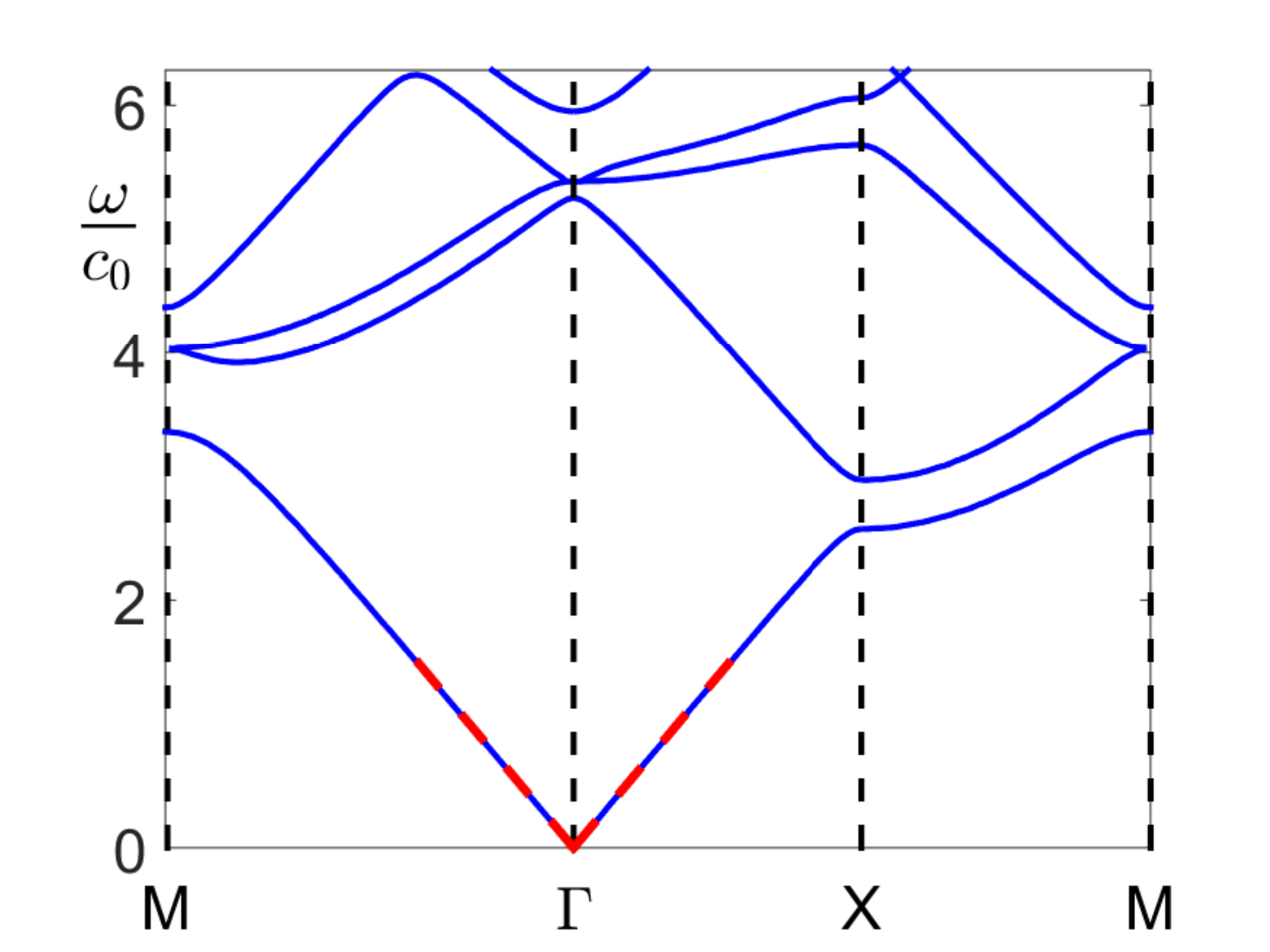}
     }
  %   \hfill
     \subfloat[\label{fig:comparebands2}]{%
       \includegraphics[width=0.3265\textwidth]{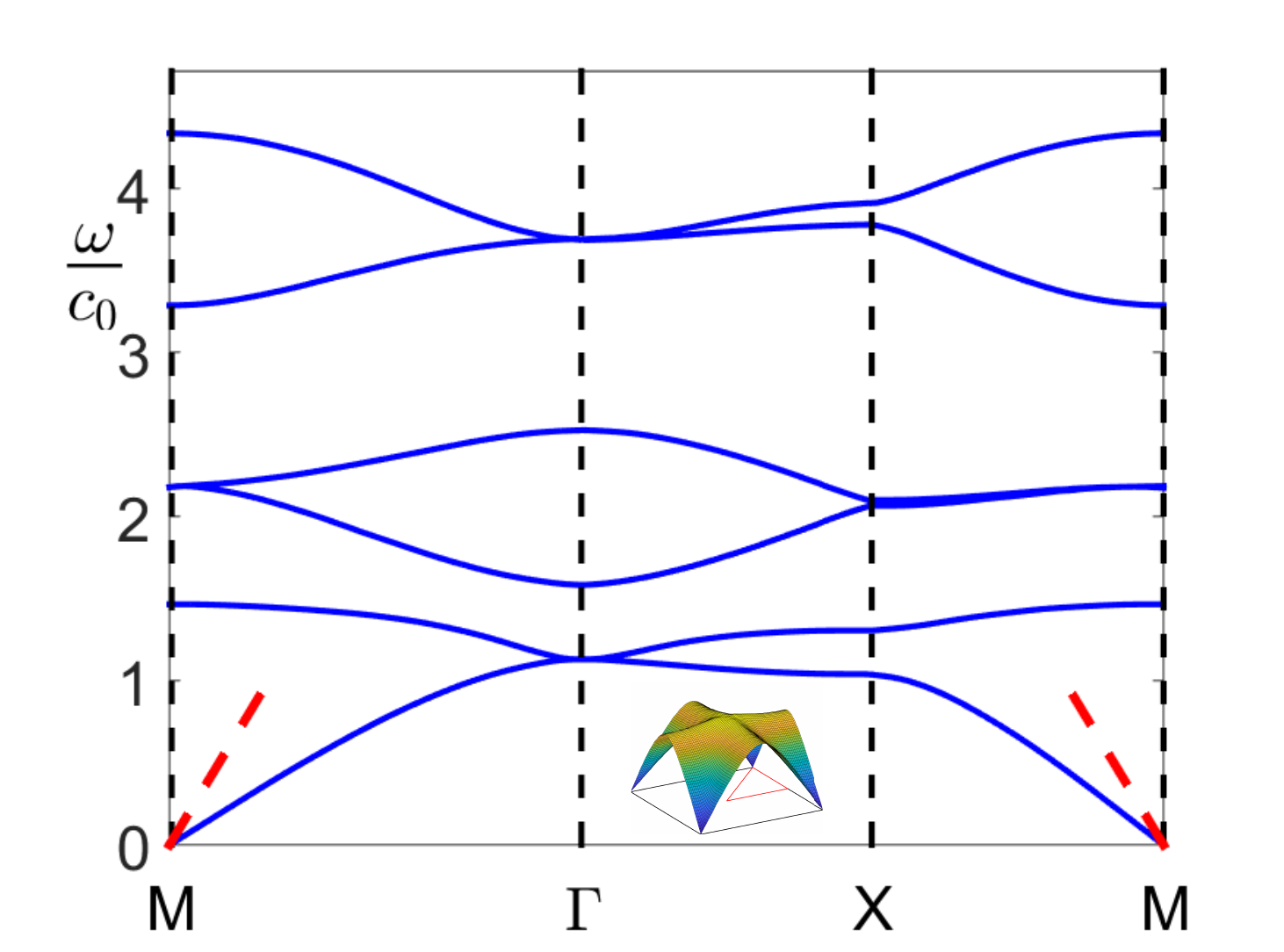}
     %  bdiag2_cts_sup fig_overlay_M_L5
     }
  \subfloat[\label{fig:comparebands3}]{%
       \includegraphics[width=0.3225\textwidth]{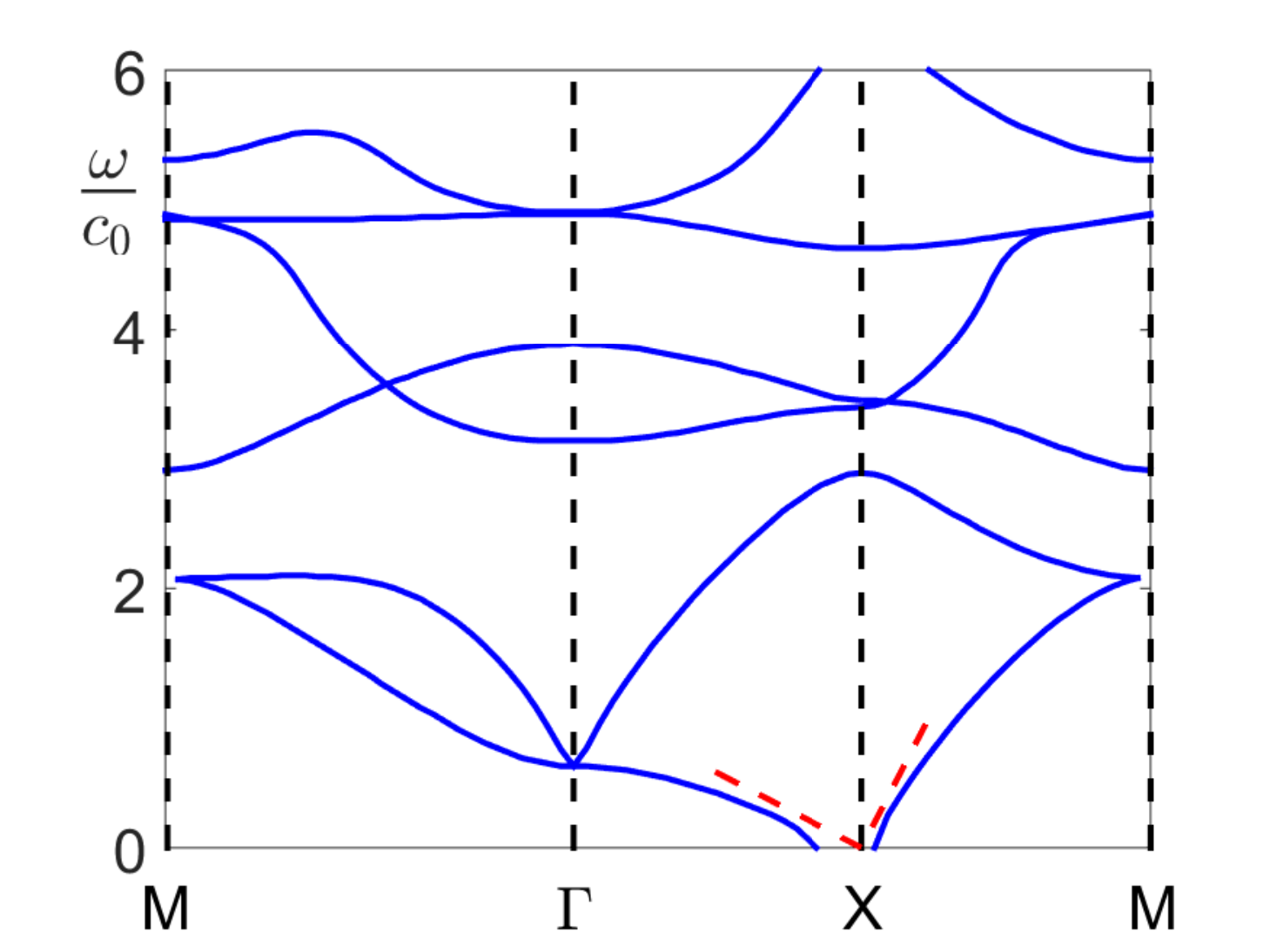}
     }

          \caption{Band diagrams  for   square array of cylinders embedded in   air   ($\varepsilon_\rmb = \mu_\rmb = 1$)  with material properties  \protect\subref{fig:comparebands1} $\varepsilon_\rmc = 1$ and $\mu_\rmc = 2$,   \protect\subref{fig:comparebands2} $\varepsilon_\rmc = -1$ and $\mu_\rmc = 2$ (inset: first band surface over first Brillouin zone), and  \protect\subref{fig:comparebands3}  $\varepsilon_\rmc \approx -0.53$ and $\mu_\rmc \approx -10.22$.  All figures use  lattice period    $a=1$,  radius   $a^\prime = 0.3a$, and truncation $L=5$.  Dashed red lines represent the dipolar first-band descriptions    \eqref{seq:disprelGamma}, \eqref{seq:kvk0}, and \eqref{seq:disprelXellipse}, respectively.  \label{fig:comparebands}   }

   \end{figure}

Returning to the discussion of anomalous resonances, these may be understood by analogy to \eqref{seq:antipov}; the zeros of $N_l$ and $N_m$ correspond to poles in the system at the  angular orders $l$ and $m$, which at low frequencies, accumulate about $\varepsilon_\rmc =-\varepsilon_\rmb$  with increasing truncation $L$ (hence why anomalous resonances are also known as accumulation points). As such, in the vicinity of  $\varepsilon_\rmc =-\varepsilon_\rmb$, the band diagrams for $L=1$ do  not necessarily reflect the band diagrams for the full system, and we have found that results using other numerical tools, such as plane-wave expansion methods \cite{busch1998photonic} and finite-element methods, exhibit {\it extremely strong} numerical instability with variations in the number of plane waves and the maximum  element size, respectively.   Preliminary results suggest that as we increase the truncation parameter $L \rightarrow \infty$, the slope of the first band  tends to $\alpha^\prime \rightarrow 0$. We also find that   the first band is only present at truncation values $L=1,2,5,6,9,10,\ldots$ (OEIS A042963) otherwise a low-frequency gap is observed; it is possible that the truncation order is tied to the existence of the first band   (i.e., the truncation must be chosen to ensure that the system possesses the correct   symmetries).  This was observed earlier, where $M$ point emanation was not supported within a monopolar truncation.   We remark that the issues observed with the $M$ point are not overcome  through the use of other solution procedures, such as  finite-element method (FEM) solvers (i.e.,   Comsol).

    In fact, conventional FEM solvers are unable to validate the bands in Fig.~1 (Letter) centered about $M$ or $X$ at low frequencies. For the $M$ point, this is because the   band diagrams returned by FEM solvers exhibit   strong instability near the anomalous resonance condition. For the $X$ point,  most commercially available solvers   cannot evaluate bands when   both the permittivity and permeability are negative-valued. However,  it is possible to examine all band diagrams in Fig.~3 (Letter) using FEM solvers,  excluding Figs.~3c and 3d, which cannot be validated due to their closeness to the anomalous resonance condition. We find that   the  conditions we derive herein give an excellent approximation to  those obtained using a full FEM solver approach  (i.e., the condition for Fig.~3e (Letter) is $\varepsilon_\rmc \approx -1.9$ using FEM solvers and not $\varepsilon_\rmc \approx -1.896$). Overall, we find  that the dipole approximation gives a good   description of both the conditions and the first band(s) when compared to those obtained with FEM solvers. For reference, the anomalous resonance condition $\varepsilon_\rmc = -\varepsilon_\rmb$ is identical for   square and hexagonal lattices of cylinders   \cite{mcphedran1980electrostatic}, and   that   no analogy was found for the hexagonal lattice under the $X$ point conditions for a square lattice.

Regarding the issue of loss, if we incorporate loss in the background material (i.e., consider complex $\varepsilon_\rmb$ and $\mu_\rmb$), and then use the  vanishing-determinant conditions to calculate  the optical properties of the cylinder   (thus obtaining complex  $\varepsilon_\rmc$ and $\mu_\rmc$), then the corresponding band diagram (under the complex $\omega$ and real $k$ representation) is robust to moderately large imaginary values. Satisfying only the real parts of the background or cylinder materials constants   may cause a  gap to emerge about a band origin   at low frequencies.  Determining materials which satisfy the vanishing determinant conditions for complex-valued constants  may pose something of a challenge, however it should be possible to engineer values for the permittivity and permeability with  metamaterials as the cylinder media, for example.

  As a final comment, we remark that   large condition numbers are intrinsic   for our cylindrical-Mie theory method, i.e. the discretised version of \eqref{seq:dispeq} or \eqref{seq:antipov}, even for conventional photonic crystals. Such  numerical ill-conditioning poses a challenge  for accurately determining   band structures   (as a numerically singular matrix  may return   spurious band surfaces). For $L=1$, we find that   condition numbers are not prohibitively  large, but due to the large values taken by $\scatcoeff_l$ and $S_l^\rmY$ as $L\rightarrow \infty$, that ill-conditioning may pose a significant challenge for larger truncations (and is not corrected by the use of \eqref{seq:antipov} in place of  \eqref{seq:dispeq}).

\clearpage
%\appendix

 \section{ Asymptotic forms of  lattice sums for a square array }
\label{sec:appendixA} \noindent
The   cylindrical-Mie  solution for a two-dimensional lattice of cylinders embedded in a uniform background material with permittivity $\varepsilon_\rmb$ and permeability $\mu_\rmb$ features      sums  of the form \cite{chin1994greens,mcphedran1996low,movchan2002asymptotic}
\begin{equation}
\label{seq:appeq1}
S_l^\rmY(\omegand,\bfk_\rmB;n_\rmb) =\sideset{}{'}\sum_{(m,n)\in\mathbb{Z}^2}  Y_l ( n_\rmb \omegand R_p) \rme^{\rmi l \phi_p} \rme^{\rmi \bfk_\rmB \cdot \bfR_p},
\end{equation}
where $Y_l(z)$ is a Bessel function  of the second kind, $\omegand = \omegad/c_0$ is the scaled angular frequency, $n_\rmb = \sqrt{\varepsilon_\rmb} \sqrt{\mu_\rmb}$, $\bfk_\rmB$ is the in-plane Bloch vector, $(R_p, \phi_p)$ represents polar coordinates for the real lattice generator $\bfR_p = (ma,na)$, for $m,n\in\mathbb{Z}$, and prime notation denotes summation over the entire lattice excluding  $(m,n)=(0,0)$, where the summand is singular. Here, the subscript $p$ represents a double index ranging over all $m$ and $n$.  The lattice sum above is conditionally convergent, however, numerous procedures exist for  accelerating convergence \cite{linton2010lattice}. One of these methods is to consider the reciprocal-space representation of \eqref{seq:appeq1}; this is readily obtained  by comparing forms of  the quasi-periodic Green's function for the Helmholtz operator defined by
\begin{equation}
\left\{ \Delta + (n_\rmb  \omegand)^2 \right\} G(\bfx , \bfx^\prime) = -2\pi \sum_{p} \delta(\bfx - \bfx^\prime - \bfR_p) \rme^{\rmi \bfk_\rmB \cdot \bfR_p}.  
\end{equation}
The comparison procedure is outlined in \citet{movchan2002asymptotic} and gives the absolutely convergent representation 
 \begin{equation}
 \label{seq:SlYapprecip}
S_l^\rmY(\omegand,\bfk_\rmB;n_\rmb ) J_l( n_\rmb \omegand\xi)  = -Y_0(n_\rmb  \omegand\xi) \delta_{l0}  
- \frac{4\rmi^l}{A} \sum_h \frac{J_l(Q_h \xi) \rme^{\rmi l \Theta_h}}{Q_h^2 - (n_\rmb\omegand)^2},  
 \end{equation}
where $A$ denotes the area of the fundamental unit cell, $(K_h,\psi_h)$ are polar coordinates for the reciprocal lattice generator $  \bfK_h =  2\pi (m/a,n/a)$ for $ m,n \in \mathbb{Z}$, $(Q_h,\Theta_h)$ denote polar coordinates for $\bfQ_h = \bfK_h + \bfk_\rmB$, and $(\xi,\gamma)$ represent  polar coordinates for $\boldsymbol{\xi} = \bfx - \bfx^\prime$ (where we remark that $\psi_h$ and $\gamma$ are unused in the above, but are defined for completeness). Here, the subscript $h$ is a double index ranging over all $m$ and $n$. Note that $\boldsymbol{\xi}$, whilst defined as the difference between the source and field coordinates for the purposes of evaluating the Green's function $G(\bfx,\bfx^\prime)$, now represents an arbitrary vector of finite length in the unit cell, as the lattice sums $S_l^\rmY$ in \eqref{seq:appeq1} are independent of spatial coordinates.

 \subsection{Asymptotic representations of dynamic lattice sums $S_l^\rmY$ near $\Gamma$ } \label{sec:SlYGpta} \noindent
 We now present a brief summary of the method for determining the asymptotic forms of $S_l^\rmY$ in the low frequency and vanishing Bloch vector limit, following the approach outlined  in Appendix A of  \citet{mcphedran1996low}. This  begins  by isolating the $h = (0,0)$ term in   \eqref{seq:SlYapprecip}, and applying Graf's addition theorem \cite[Eq. (9.1.79)]{abramowitz1964handbook} to the surviving summand, admitting 
 \begin{multline}
 \label{seq:SlYgraf}
 S_l^\rmY(\omegand,\bfk_\rmB;n_\rmb)   = - \frac{Y_0(n_\rmb\omegand\xi)}{J_0(n_\rmb\omegand\xi)} \delta_{l0} - \frac{4\rmi^l}{A } \frac{J_l(k_\rmB \xi)}{J_l(n_\rmb\omegand\xi)}  \frac{\rme^{\rmi l \theta_\rmB}}{ k_\rmB^2 - (n_\rmb\omegand)^2} \\
 - \frac{4\rmi^l}{A } \frac{1}{J_l(n_\rmb\omegand\xi)} \sideset{}{'}\sum_{h} \sum_{m=-\infty}^{\infty} \frac{1}{Q_h^2 - (n_\rmb\omegand)^2} \left\{ (-1)^m J_{l+m} (K_h \xi)   J_m(k_\rmB \xi) \rme^{\rmi (l+m) \psi_h - \rmi m \theta_\rmB} \right\},
 \end{multline} 
 where $(k_\rmB,\theta_\rmB)$ denotes the polar representation of the Bloch vector $\bfk_\rmB$. A Taylor series expansion for small $\omegand$ and $k_\rmB$ gives
 \begin{equation}
 \label{seq:Qhlim}
 \lim_{k_\rmB \rightarrow 0 } \lim_{\omegand \rightarrow 0}  \frac{1}{Q_h^2 - (n_\rmb\omegand)^2} \approx \frac{1}{K_h^2} \left(  1 -   \frac{2 k_\rmB}{K_h} \cos(\psi_h - \theta_\rmB)  + \frac{2 k_\rmB^2}{K_h^2} \cos(2\left\{\psi_h - \theta_\rmB \right\} ) + \frac{k_\rmB^2 +(n_\rmb\omegand)^2}{K_h^2} + \ldots \right),
 \end{equation}
 which after an appropriate truncation of $m$ in \eqref{seq:SlYgraf}, and considerable algebraic manipulation, admits the asymptotic representations  
 \begin{subequations}
\label{seq:allSlYinitG}
\begin{align}
\lim_{k_\rmB \rightarrow 0} \lim_{\omegand \rightarrow 0} \left\{ S_0^Y \right\} &\approx - \frac{Y_0(n_\rmb\omegand\xi)}{J_0(n_\rmb\omegand\xi)} - \frac{4}{A} \left\{   \left( \frac{1}{k_\rmB^2 - (n_\rmb\omegand)^2} + S_{0,0,2} \right)  \,  \frac{J_0(k_\rmB \xi)}{J_0(n_\rmb\omegand\xi)} + 2 k_\rmB  S_{1,0,3}    \frac{J_1(k_\rmB  \xi)}{J_0(n_\rmb\omegand\xi)}  + \ldots \right\},\\
\lim_{k_\rmB \rightarrow 0} \lim_{\omegand \rightarrow 0} \left\{ S_1^Y \right\} &\approx  - \frac{4\rmi \rme^{\rmi \theta_\rmB }}{A} \left\{   \left( \frac{1}{k_\rmB^2 - (n_\rmb\omegand)^2} + S_{0,0,2} \right)   \,  \frac{J_1(k_\rmB \xi)}{J_1(n_\rmb\omegand\xi)} - k_\rmB  S_{1,0,3}    \frac{J_0(k_\rmB  \xi)}{J_1(n_\rmb\omegand\xi)}+ \ldots \right\}, \\ \nonumber
\lim_{k_\rmB \rightarrow 0} \lim_{\omegand \rightarrow 0} \left\{ S_2^Y \right\}  &\approx \frac{4 \rme^{2 \rmi \theta_\rmB } }{A} \left\{ \left( \frac{1}{k_\rmB^2 - (n_\rmb\omegand)^2} + S_{0,0,2} \right)    \frac{J_2(k_\rmB \xi)}{J_2(n_\rmb\omegand\xi)} - k_\rmB  \, S_{1,0,3}  \frac{J_1(k_\rmB \xi)}{J_2(n_\rmb\omegand\xi)}   + k_\rmB^{\prime 2} \,  S_{2,0,4}   \frac{J_0(k_\rmB  \xi)}{J_2(n_\rmb\omegand\xi)}   \right\} \\
&\qquad +\frac{4 \rme^{-2 \rmi \theta_\rmB } }{A} \left\{  S_{4,4,2}  \, \frac{J_2(k_\rmB \xi)}{J_2(n_\rmb\omegand\xi)}  + k_\rmB  \, S_{3,4,3}  \, \frac{J_1(k_\rmB \xi)}{J_2(n_\rmb\omegand\xi)}    +k_\rmB^{  2} \, S_{2,4,4}  \,   \frac{J_0(k_\rmB \xi)}{J_2(n_\rmb\omegand\xi)}   + \ldots \right\},
\end{align}
\end{subequations}
 where we define the double Sch\"{o}milch series \cite{chen2016evaluation}
 \begin{equation}
 \label{seq:cyllattG}
 S_{l,m,n}  (\xi;\tau,a)= \sum_{h} {}^\prime \frac{J_l(K_h \xi)}{K_h^{n}} \rme^{\rmi m \psi_h}.
 \end{equation}
Using the explicit representations for $S_{l,m,n}$ in \citet{chen2016evaluation}, or the recurrence relation procedure in  \citet{nicorovici1996analytical,mcphedran1996low,chen2016evaluation}, we obtain    
 
\begin{subequations}
 \begin{align}
S_{0,0,2}  &= - \frac{a^2}{2\pi} \log(\xi) + \frac{a^2}{4\pi} \log\left(\frac{4 \pi  a^2 }{\Gamma(\tfrac{1}{4})^4}\right) +\frac{\xi^2}{4}, \quad &&
  S_{2,4,4}   =  \frac{a^2 \xi^2 \Gamma(\tfrac{1}{4})^8}{3 \pi^5 2^{12}}    -   \frac{\xi^4 \Gamma(\tfrac{1}{4})^8}{9 \pi^4 2^{11}}  +    \frac{\xi^6 \Gamma(\tfrac{1}{4})^8}{15 \pi^3 a^2 2^{13} }, \\
  S_{1,0,3} &= - \frac{a^2 \xi}{4\pi} \log(\xi) + \frac{a^2 \xi}{8\pi} \log\left( \frac{4 \pi a^2  }{\Gamma(\tfrac{1}{4})^4}\right) + \frac{a^2 \xi}{8\pi} + \frac{\xi^3}{16}, \quad  &&
S_{3,4,3}   =  \frac{\xi^3\Gamma(\tfrac{1}{4})^8}{9 \pi^4 2^{10}}   -   \frac{\xi^5 \Gamma(\tfrac{1}{4})^8}{15 \pi^3 a^2  2^{11}},\\
S_{2,0,4}  &= - \frac{a^2 \xi^2}{16\pi} \log\left(\xi\right) + \frac{a^2 \xi^2}{8\pi} \log\left(\frac{4 \pi a^2  }{\Gamma(\tfrac{1}{4})^4}\right)  + \frac{3a^2 \xi^2}{64\pi} + \frac{\xi^4}{96}, \quad    &&
 S_{4,4,2}   =   \frac{\xi^4  \Gamma(\tfrac{1}{4})^8}{15 \pi^3 a^2 2^{10}} ,
  \end{align}
  \end{subequations}
  for a square lattice of period $a$. The  $S_{l,m,n}$ expressions above in tandem with the small argument expansions      \cite{mcphedran1996low}  
\begin{subequations}
\label{seq:allBesselexp}
\begin{align}
\lim_{\xi \rightarrow 0} \lim_{k_\rmB \rightarrow 0} \lim_{\omegand \rightarrow 0} \frac{Y_0(n_\rmb\omegand\xi)}{J_0(n_\rmb\omegand\xi)} &\approx \frac{2}{\pi}\left( \gamma_\rme + \log\left( \frac{n_\rmb\omegand\xi}{2}\right) \right)+ \frac{(n_\rmb\omegand)^2 \xi^2}{2 \pi} + \ldots, \\
\lim_{\xi \rightarrow 0} \lim_{k_\rmB \rightarrow 0} \lim_{\omegand \rightarrow 0}\frac{J_0(k_\rmB\xi)}{J_0(n_\rmb\omegand\xi)} &\approx 1 + \frac{(n_\rmb\omegand)^2 - k_\rmB^2}{4} \xi^2 + \ldots, \\
\lim_{\xi \rightarrow 0} \lim_{k_\rmB \rightarrow 0} \lim_{\omegand \rightarrow 0}\frac{J_1(k_\rmB\xi)}{J_1(n_\rmb\omegand\xi)} &\approx \frac{k_\rmB}{n_\rmb\omegand} + \frac{k_\rmB}{n_\rmb\omegand} \frac{(n_\rmb\omegand)^2 - k_\rmB^2}{8} \xi^2 + \ldots, \\
\lim_{\xi \rightarrow 0} \lim_{k_\rmB \rightarrow 0} \lim_{\omegand \rightarrow 0}\frac{J_0(k_\rmB\xi)}{J_1(n_\rmb\omegand\xi)} &\approx  \frac{2}{n_\rmb\omegand \xi} + \frac{(n_\rmb\omegand)^2 - 2k_\rmB^2}{4 n_\rmb\omegand} \xi + \left( \frac{1}{32} \frac{k_\rmB}{n_\rmb\omegand} - \frac{1}{16} \frac{n_\rmb\omegand}{k_\rmB} + \frac{1}{48} \frac{(n_\rmb\omegand)^3}{k_\rmB^3} \right) k_\rmB^3 \xi^3 + \ldots, \\
\lim_{\xi \rightarrow 0} \lim_{k_\rmB \rightarrow 0} \lim_{\omegand \rightarrow 0}\frac{J_1(k_\rmB\xi)}{J_0(n_\rmb\omegand\xi)} &\approx \frac{k_\rmB \xi }{2} + \frac{2(n_\rmb\omegand)^2 - k_\rmB^2}{16} k_\rmB \xi^3 + \ldots, \\
 \lim_{\xi \rightarrow 0} \lim_{k_\rmB \rightarrow 0} \lim_{\omegand \rightarrow 0}\frac{J_0(k_\rmB\xi)}{J_2(n_\rmb\omegand\xi)} &\approx \frac{8}{(n_\rmb\omegand)^2 \xi^2} + 2\left( \frac{1}{3} - \frac{k_\rmB^2}{(n_\rmb\omegand)^2} \right) + \left( -\frac{1}{6} + \frac{1}{8}\frac{k_\rmB^2}{ (n_\rmb\omegand)^2} + \frac{5}{144} \frac{(n_\rmb\omegand)^2}{k_\rmB^2} \right) k_\rmB^2 \xi^2 + \ldots, \\
\lim_{\xi \rightarrow 0} \lim_{k_\rmB \rightarrow 0} \lim_{\omegand \rightarrow 0}\frac{J_1(k_\rmB\xi)}{J_2(n_\rmb\omegand\xi)} &\approx  \frac{4 k_\rmB}{(n_\rmb\omegand)^2\xi} + \frac{2(n_\rmb\omegand)^2 - 3k_\rmB^2}{6(n_\rmb\omegand)^2} k_\rmB \xi + \left( -\frac{1}{24} + \frac{1}{48} \frac{k_\rmB^2}{(n_\rmb\omegand)^2} + \frac{5}{288} \frac{(n_\rmb\omegand)^2}{k_\rmB^2} \right) k_\rmB^3 \xi^3 + \ldots, \\
\lim_{\xi \rightarrow 0} \lim_{k_\rmB \rightarrow 0} \lim_{\omegand \rightarrow 0}\frac{J_2(k_\rmB\xi)}{J_2(n_\rmb\omegand\xi)} &\approx \frac{k_\rmB^2}{(n_\rmb\omegand)^2} +\frac{k_\rmB^2}{(n_\rmb\omegand)^2} \frac{(n_\rmb\omegand)^2 - k_\rmB^2}{12}\xi^2 +\ldots,
\end{align}
\end{subequations}
  finally admit the representations \eqref{seq:asySlYGamma} for $S_0^\rmY$, $S_1^\rmY$, and $S_2^\rmY$ both at low frequencies and in the vicinity of the $\Gamma$ point, after extensive algebraic manipulation. 
 \subsection{Asymptotic representations of dynamic lattice sums $S_l^\rmY$ near $M$ } \label{sec:SlYMpta} \noindent
To obtain closed-form representations for  $S_l^\rmY$ in the vicinity of  different symmetry points (and also for vanishing $\omegand$) we extend  the procedure   above. The approach is identical up to \eqref{seq:SlYgraf}, however, to consider behaviour near the $M$ point it is necessary to modify   \eqref{seq:Qhlim} appropriately.  This is achieved by decomposing  the translated reciprocal lattice generator as $\bfQ_h = \bfK_h + \bfk_\rmB = (\bfK_h + \bfM)+ (\bfk_\rmB - \bfM)  = \bfK_h^\prime + \bfk_\rmB^\prime$, where $\bfM = (\pi/a,\pi/a)$. Substituting this decomposition into \eqref{seq:Qhlim}, we recover an identical    expression   to before, but with the replacements  $\bfK_h \mapsto \bfK_h^\prime$ and $\bfk_\rmB \mapsto \bfk_\rmB^\prime$    and an updated   limit argument. Subsequently we obtain analogous expressions to \eqref{seq:allSlYinitG} but without the  singular terms $1/(k_\rmB^2 - (n_\rmb\omegand)^2)$, as   light lines are not present near   $M$   at low frequencies. That is, for the $M$ point we obtain
\begin{subequations}
\label{seq:allSlYinit}
\begin{align}
\lim_{k_\rmB \rightarrow M} \lim_{ \omegand \rightarrow 0} \left\{ S_0^Y \right\} &\approx - \frac{Y_0(n_\rmb\omegand\xi)}{J_0(n_\rmb\omegand\xi)} - \frac{4}{A} \left\{   S_{0,0,2}^\rmM  \,  \frac{J_0(k_\rmB^\prime \xi)}{J_0(n_\rmb\omegand\xi)} + 2 k_\rmB^\prime  S_{1,0,3}^\rmM   \frac{J_1(k_\rmB^\prime  \xi)}{J_0(n_\rmb\omegand\xi)} \right\},\\
\lim_{k_\rmB \rightarrow M} \lim_{ \omegand \rightarrow 0} \left\{ S_1^Y \right\} &\approx  - \frac{4\rmi \rme^{\rmi \theta_\rmB^\prime }}{A} \left\{   S_{0,0,2}^\rmM  \,  \frac{J_1(k_\rmB^\prime \xi)}{J_1(n_\rmb\omegand\xi)} - k_\rmB^\prime  S_{1,0,3}^\rmM   \frac{J_0(k_\rmB^\prime  \xi)}{J_1(n_\rmb\omegand\xi)} \right\}, \\ \nonumber
\lim_{k_\rmB \rightarrow M} \lim_{ \omegand \rightarrow 0} \left\{ S_2^Y \right\}  &\approx \frac{4 \rme^{2 \rmi \theta_\rmB^\prime } }{A} \left\{  S_{0,0,2}^\rmM   \frac{J_2(k_\rmB^\prime \xi)}{J_2(n_\rmb\omegand\xi)} - k_\rmB^\prime  \, S_{1,0,3}^\rmM \frac{J_1(k_\rmB^\prime \xi)}{J_2(n_\rmb\omegand\xi)}   + k_\rmB^{\prime 2} \,  S_{2,0,4}^\rmM  \frac{J_0(k_\rmB^\prime  \xi)}{J_2(n_\rmb\omegand\xi)}   \right\} \\
&\qquad +\frac{4 \rme^{-2 \rmi \theta_\rmB^\prime } }{A} \left\{  S_{4,4,2}^\rmM \, \frac{J_2(k_\rmB^\prime \xi)}{J_2(n_\rmb\omegand\xi)}  + k_\rmB^\prime  \, S_{3,4,3}^\rmM \, \frac{J_1(k_\rmB^\prime \xi)}{J_2(n_\rmb\omegand\xi)}    +k_\rmB^{\prime 2} \, S_{2,4,4}^\rmM \,   \frac{J_0(k_\rmB^\prime \xi)}{J_2(n_\rmb\omegand\xi)}    \right\},
\end{align}
\end{subequations}
where we emphasise that  $\boldsymbol{\xi}$ is an arbitrary vector in the first Brillouin zone that remains untranslated, $(k_\rmB^\prime ,\theta_\rmB^\prime )$ is the polar representation of $\bfk_\rmB^\prime = \bfk_\rmB - \bfM$, and  
\begin{equation}
S_{l,m,n}^\rmM (\xi;\tau,a)= \sum_{h} {}^\prime \frac{J_l(K_h^\prime \xi)}{K_h^{\prime n}} \rme^{\rmi m \psi_h^\prime},
\end{equation}
where $(K_h^\prime ,\psi_h^\prime )$ is the polar representation of $\bfK_h^\prime = \bfK_h + \bfM$.  As highlighted in  \citet{chen2016evaluation}, analytical expressions for  $S_{l,m,n}^\rmM$, corresponding to      Bloch vectors centred around the $M$ point, are easily obtained by evaluating   a selection of  rectangular and square $\Gamma$-centred sums and using  the multi-set identity \cite{chen2016evaluation}
 \begin{subequations}
\begin{equation}
\label{seq:multisetM}
\overbar{\Omega}^M(\rmi,a) = \overbar{\Omega}(\rmi,2a) - \overbar{\Omega}(\rmi/2,2a) + \overbar{\Omega}(\rmi,a) - \overbar{\Omega}(2\rmi,a),
\end{equation}
where $\overbar{\Omega}$ are  reciprocal lattice sets defined by
\begin{equation}
\overbar{\Omega}(\tau,a) = \left\{ \frac{2\pi}{A} (h_1 b \hat{\bfe}_1^\prime + h_2 a \hat{\bfe}_2^\prime )  \bigg| h_1,h_2 \in \mathbb{Z}\right\},
\end{equation}
 \end{subequations}
with $A = ab $ representing the area of the unit cell (for a rectangular lattice of periods $a$ and $b$ in $x$ and $y$ respectively), $\hat{\bfe}_j^\prime$ are   basis vectors for the reciprocal lattice,   and $\tau = (b/a) \rmi$.  The   multi-set identity \eqref{seq:multisetM} gives the following expressions
\begin{subequations}
\label{seq:allSlmnM}
\begin{align}
S_{0,0,2}^\rmM  &= - \frac{a^2}{2 \pi } \log \left( \xi  \right)  + \frac{a^2}{4 \pi }  \log  \left( \frac{ 16 \pi a^2 }{\Gamma  (\tfrac{1}{4} )^{4} } \right) ,  \quad
  &&S_{2,4,4}^\rmM  =   - \frac{\xi^6 \Gamma(\tfrac{1}{4} )^8 }{5 \pi^3 a^2  2^{14}} + \frac{ \xi^4 \Gamma(\tfrac{1}{4} )^8 }{9 \pi^4 2^{10}}   -\frac{a^2 \xi^2 \Gamma(\tfrac{1}{4} )^8  }{  \pi^5 2^{12}} ,  \\
  S_{1,0,3}^\rmM   &=  \frac{a^2 \xi}{8 \pi } -  \frac{a^2 \xi}{4 \pi }   \log \left( \xi \right) +  \frac{a^2 \xi}{8 \pi }  \log \left( \frac{ 16 \pi a^2 }{\Gamma  (\tfrac{1}{4} )^{4} } \right) ,\quad
&&S_{3,4,3}^\rmM  =    \frac{ \xi^5 \Gamma(\tfrac{1}{4} )^8 }{5  \pi^3 a^2 2^{12} }  -\frac{ \xi^3 \Gamma(\tfrac{1}{4})^8 }{9 \pi^4 2^9 },  \\ 
S_{2,0,4}^\rmM  &=   \frac{3 a^2 \xi^2 }{64 \pi }  - \frac{a^2 \xi^2 }{16 \pi }  \log \left(\xi \right) + \frac{a^2 \xi^2 }{32 \pi } \log  \left( \frac{ 16 \pi a^2 }{\Gamma  (\tfrac{1}{4} )^{4} } \right)  ,  \quad 
&&S_{4,4,2}^\rmM   =  -\frac{  \xi^4 \Gamma(\tfrac{1}{4} )^8 }{5  \pi^3 a^2 2^{11}  }.
\end{align}
\end{subequations}
Substituting the $S_{l,m,n}^\rmM$ expressions   \eqref{seq:allSlmnM} and the Bessel function expansions \eqref{seq:allBesselexp} into the expansions for $S_l^\rmY$   \eqref{seq:allSlYinit} above  ultimately gives the final expressions \eqref{seq:SmY} used in Section \ref{sec:SlYMpt}.

\subsection{Asymptotic representations of  $S_l^\rmY$ at the $X$ point}\label{seq:XpointSlmnSlY} \noindent 
As with the $M$ point approach, in order to consider behaviour near the $X$ point it is necessary to modify   \eqref{seq:Qhlim} appropriately. Decomposing  $\bfQ_h = \bfK_h + \bfk_\rmB = (\bfK_h + \bfX)+ (\bfk_\rmB - \bfX) $, where $\bfX = (\pi/a,0)$ into \eqref{seq:Qhlim}   we obtain
 \begin{subequations}
\label{seq:SlYappX}
\begin{multline}
\lim_{k_\rmB \rightarrow X}  \lim_{\omegand \rightarrow 0} 
S_0^\rmY \approx -\frac{Y_0(n_\rmb\omegand\xi)}{J_0(n_\rmb\omegand\xi)} 
 - \frac{4}{A}  \frac{J_0(k_\rmB^\pp \xi)}{J_0(n_\rmb\omegand \xi)} \left(  S_{0,0,2}^\rmX 
+ (n_\rmb\omegand)^2 S_{0,0,4}^\rmX  
 +   k_\rmB^{\pp 2} S_{0,0,4}^\rmX     \right)\\
 - \frac{4}{A}  \frac{J_0(k_\rmB^\pp \xi)}{J_0(n_\rmb\omegand \xi) }  \rme^{-2\rmi \theta_\rmB^\pp }   \left( k_\rmB^{\pp 2}  S_{0,2,4}^\rmX   
 \right) 
 - \frac{4}{A}  \frac{J_0(k_\rmB^\pp \xi)}{J_0(n_\rmb\omegand \xi)}  \rme^{2\rmi \theta_\rmB^\pp } \left(  k_\rmB^{\pp 2}  S_{0,2,4}^\rmX     \right),
\end{multline}
 \begin{multline}
\lim_{k_\rmB \rightarrow X}  \lim_{\omegand \rightarrow 0} S_1^\rmY \approx
-\frac{4\rmi \rme^{\rmi \theta_\rmB^\pp}  }{A} \frac{J_1(k_\rmB^\pp \xi)}{J_1(n_\rmb\omegand  \xi)} 
\left( S_{0,0,2}^\rmX  
+(n_\rmb\omegand)^2 S_{0,0,4}^\rmX  
+ k_\rmB^{\pp 2}  S_{0,0,4}^\rmX  
 \right) 
+\frac{4\rmi \rme^{\rmi \theta_\rmB^\pp}    }{A} \frac{J_0(k_\rmB^\pp \xi)}{J_1(n_\rmb\omegand  \xi)} \left( 
  k_\rmB^{\pp  }   S_{1,0,3}^\rmX   
 \right) \\
+\frac{4\rmi \rme^{-\rmi \theta_\rmB^\pp}   }{A} \frac{J_0(k_\rmB^\pp \xi) }{J_1(n_\rmb\omegand  \xi) }  \left( 
 k_\rmB^{\pp  }   S_{1,2,3}^\rmX
 \right)  
-\frac{4\rmi \rme^{-\rmi \theta_\rmB^\pp}    }{A}  \frac{J_1(k_\rmB^\pp \xi)}{J_1(n_\rmb\omegand  \xi)} \left( k_\rmB^{\pp 2}   S_{0,2,4}^\rmX     \right) ,
\end{multline}
 \begin{multline}
\lim_{k_\rmB \rightarrow X}  \lim_{\omegand \rightarrow 0} S_2^\rmY \approx  
 \frac{4  }{A}\frac{J_0(k_\rmB^\pp \xi)}{J_2(n_\rmb\omegand  \xi)} \left( S_{2,2,2}^\rmX + (n_\rmb\omegand)^{ 2} S_{2,2,4}^\rmX  +  k_\rmB^{\pp 2} S_{2,2,4}^\rmX   \right) 
 -\frac{4  }{A}  \frac{J_1(k_\rmB^\pp \xi)}{J_2(n_\rmb\omegand  \xi)} \left( k_\rmB^\pp S_{1,2,3}^\rmX \right) \\
+ \frac{4  }{A} \frac{J_2(k_\rmB^\pp \xi)}{J_2(n_\rmb\omegand  \xi)} \left( k_\rmB^{\pp 2} S_{0,2,4}^\rmX \right)  
+ \frac{4 \rme^{2\rmi \theta_\rmB^\pp} }{A}   \frac{J_2(k_\rmB^\pp \xi)}{J_2(n_\rmb\omegand  \xi)}  \left(  S_{0,0,2}^\rmX + (n_\rmb\omegand)^2 S_{0,0,4}^\rmX +    k_\rmB^{\pp 2} S_{0,0,4}^\rmX 
     \right) \\
     +  \frac{4 \rme^{2\rmi \theta_\rmB^\pp}  }{A}  \frac{J_1(k_\rmB^\pp \xi) }{J_2(n_\rmb\omegand  \xi) }  \left(  -  k_\rmB^\pp S_{1,0,3}^\rmX \right) 
     +   \frac{4 \rme^{2\rmi \theta_\rmB^\pp}  }{A}  \frac{J_0(k_\rmB^\pp \xi) }{J_2(n_\rmb\omegand  \xi) }  \left(k_\rmB^{\pp 2} S_{2,0,4}^\rmX  
\right) 
+\frac{4  \rme^{-2\rmi \theta_\rmB^\pp} }{A} \frac{J_0(k_\rmB^\pp \xi)}{J_2(n_\rmb\omegand  \xi)}  \left( 
 k_\rmB^{\pp 2}S_{2,4,4}^\rmX 
\right),
\end{multline}
\end{subequations}
where $\bfk_\rmB^\pp = \bfk_\rmB - \bfX$ with polar representation $(k_\rmB^\pp,\theta_\rmB^\pp)$, $\bfK_h^\pp = \bfK_h + \bfX$ with polar representation $(K_h^\pp ,\psi_h^\pp )$, and
\begin{equation}
S_{l,m,n}^\rmX (\xi;\tau,a)= \sum_{h} {}^\prime \frac{J_l(K_h^\pp \xi)}{K_h^{\pp n}} \rme^{\rmi m \psi_h^\pp},
\end{equation}
 Note that  in contrast to the representations for $S_l^\rmY$ in \eqref{seq:allSlYinitG} and \eqref{seq:allSlYinit},  the closed-form representations for $S_l^\rmY$ near the $X$ point have a considerably different    structure. This is because  only two-fold symmetry is possessed near the $X$ point at low frequencies.  The  $S_{l,m,n}^\rmX $ are evaluated using  the multi-set identity   \cite{chen2016evaluation}
\begin{equation}
\label{eq:multiX}
\overbar{\Omega}^X(\rmi,a) =  \overbar{\Omega}(\rmi/2,2a) - \overbar{\Omega}(\rmi,a)  ,
\end{equation}
to obtain
\begin{subequations}
\label{seq:SlmnXapp}
\begin{align}
S_{0,0,2}^\rmX  &=  -\frac{a^2}{2 \pi } \log (\xi )  +\frac{a^2 }{4 \pi }\log \left(\frac{    32 \pi a^2   }{   \Gamma  (\tfrac{1}{4} )^4  }   \right) ,\quad S_{0,0,4}^\rmX  =   \frac{3 a^4   }{2 \pi^4 } \zeta(2) \beta(2) -\frac{a^2 \xi^2}{8\pi} + \frac{a^2 \xi^2}{8\pi} \log(\xi) 
- \frac{a^2 \xi^2}{16 \pi} \log \left(\frac{32 \pi a^2   }{\Gamma(\tfrac{1}{4})^4} \right), \\
S_{0,2,4}^\rmX  &= -\frac{a^2 \Gamma  (\tfrac{1}{4} )^4}{128 \pi^3 }\xi^2 + \frac{\Gamma  (\tfrac{1}{4} )^4}{1024 \pi^2}\xi^4 - \frac{a^4}{16 \pi^4} G_4^{(2)}(\tfrac{\rmi}{2}),  \quad S_{1,0,3}^\rmX  =   \frac{a^2 \xi}{8\pi} - \frac{a^2 \xi}{4\pi} \log(\xi) + \frac{a^2 \xi }{8 \pi }\log \left(\frac{ 32 \pi a^2 }{     \Gamma  (\tfrac{1}{4} )^4     }   \right), \\
S_{1,2,3}^\rmX   &= \frac{a^2 \Gamma  (\tfrac{1}{4} )^4  }{64 \pi^3} \xi - \frac{\Gamma  (\tfrac{1}{4} )^4  }{256 \pi^2} \xi^3, \quad S_{2,0,4}^\rmX  =   \frac{3 a^2 \xi^2}{64\pi} - \frac{a^2 \xi^2}{16\pi} \log(\xi)  + \frac{a^2 \xi^2}{32 \pi} \log \left( \frac{32  \pi a^2 }{  \Gamma(\tfrac{1}{4})^4}\right), \\
S_{2,2,2}^\rmX   &=  \frac{\Gamma  (\tfrac{1}{4} )^4}{128 \pi^2} \xi^2,  \quad
S_{2,2,4}^\rmX   =  \frac{a^2 \Gamma  (\tfrac{1}{4} )^4}{256 \pi^3} \xi^2  - \frac{\Gamma  (\tfrac{1}{4} )^4}{3 \pi^2 2^9} \xi^4, \quad
S_{2,4,4}^\rmX  =\frac{\xi^6 \Gamma(\tfrac{1}{4})^8 }{5 \pi^3 a^2 2^{16}}  - \frac{\xi^4 \Gamma(\tfrac{1}{4})^8}{9 \pi^4 2^{11}} + \frac{a^2 \xi^2 \Gamma(\tfrac{1}{4})^8}{  \pi^5 2^{12}},  
\end{align}
\end{subequations}
where $  \zeta(s)$ denotes the Riemann zeta function and $ \beta(s)$ the Dirichlet beta function. 
 Note that closed form expressions are not generally available for phase-conjugated Eisenstein series $G_n^{(m)}$ when $m<n$ (for reference,  $G_4^{(2)}(\tfrac{\rmi}{2}) \approx -30.854212880849047$ and is given as a rapidly convergent sum in \citet{chen2016evaluation}). 
Substituting  \eqref{seq:allBesselexp} and \eqref{seq:SlmnXapp} into the asymptotic approximations \eqref{seq:SlYappX} above gives the final expressions \eqref{seq:asySlYX} presented in Section \ref{sec:SlYXpt}.

As a final comment, we remark that obtaining closed-form expressions   for $S_l^\rmY$ about   other points in the Brillouin zone may pose something of a challenge as these do not have multiset representations using origin centered lattices. The $X$ and $M$ points are special cases where multiset representations are possible due to symmetry, despite not being origin centered lattices themselves.

 \bibliography{bibmerged}

\end{document}